\documentstyle[12pt,aaspp4]{article}

\newbox\grsign \setbox\grsign=\hbox{$>$} \newdimen\grdimen \grdimen=\ht\grsign
\newbox\simlessbox \newbox\simgreatbox
\setbox\simgreatbox=\hbox{\raise.5ex\hbox{$>$}\llap
     {\lower.5ex\hbox{$\sim$}}}\ht1=\grdimen\dp1=0pt
\setbox\simlessbox=\hbox{\raise.5ex\hbox{$<$}\llap
     {\lower.5ex\hbox{$\sim$}}}\ht2=\grdimen\dp2=0pt
\def\simgreat{\mathrel{\copy\simgreatbox}}
\def\simless{\mathrel{\copy\simlessbox}}

\newcommand{\etal}{{et al.}\ }

\newcommand{\as}{^{\prime\prime}}
\newcommand{\nlarge}{256}
\newcommand{\nxray}{101}

\newcommand{\ndeep}{20}
\newcommand{\ndeepx}{12}

\newcommand{\uniquepairs}{105}
\newcommand{\nredtot}{877}
\newcommand{\nredold}{154}

\newcommand{\nfit}{738}

\newcommand{\uniquediss}{41}

\newcommand{\m}{$^{-1}$}
\newcommand{\vbar}{\langle v \rangle}
\newcommand{\fka}{f_{ka}}
\newcommand{\Var}{{\mathrm Var}}

\newcommand{\degr}{$^\circ$}
\newcommand{\rCC}{r_{200}}
\newcommand{\Rt}{\tilde{R}}
\newcommand{\vrthree}{\overline{v_r^2}}
\newcommand{\vrbar}{\tilde{v}}
\newcommand{\vtthree}{\overline{v_\theta^2}}
\newcommand{\rt}{\tilde{r}}
\newcommand{\rtpone}{\left( \rt + 1 \right)}

\begin{document}

\title{The Dynamics of Poor Systems of Galaxies}

\author{Andisheh Mahdavi and Margaret J. Geller}
\authoremail{amahdavi@cfa.harvard.edu; mgeller@cfa.harvard.edu}
\affil{Harvard-Smithsonian Center for Astrophysics}

\author{Hans B\"ohringer}
\authoremail{hxb@rosat.mpe-garching.mpg.de}
\affil{Max-Planck-Institut f\"ur Extraterrestrische Physik}

\author{Michael J. Kurtz}
\authoremail{mkurtz@cfa.harvard.edu}
\affil{Harvard-Smithsonian Center for Astrophysics}

\and

\author{Massimo Ramella}
\authoremail{ramella@oat.ts.astro.it}
\affil{Osservatorio Astronomico di Trieste}

\clearpage

\begin{abstract}

We assemble and observe a sample of poor galaxy systems that is
suitable for testing N-body simulations of hierarchical clustering
(Navarro, Frenk, \& White 1997; NFW) and other dynamical halo models
(e.g., Hernquist 1990). We (1) determine the parameters of the density
profile $\rho(r)$ and the velocity dispersion profile $\sigma_p(R)$,
(2) separate emission-line galaxies from absorption-line galaxies,
examining the model parameters and as a function of spectroscopic
type, and (3) for the best-behaved subsample, constrain the velocity
anisotropy parameter, $\beta$, which determines the shapes of the
galaxy orbits.

Our sample consists of \ndeep\ systems, \ndeepx\ of which have
extended x-ray emission in the ROSAT All-Sky Survey. We measure the
\nredtot\ optical spectra of galaxies brighter than $m_R \approx 15.4$
within 1.5$h$\m\ Mpc of the system centers (we take $H_0 = 100h$ km
s\m\ Mpc\m). Thus we sample the system membership to a radius typically
three times larger than other recent optical group surveys. The
average system population is $30$ galaxies, and the average
line-of-sight velocity dispersion is $\approx 300$ km s\m.

The NFW universal profile and the Hernquist (1990) model both provide
good descriptions of the spatial data.  In most cases an isothermal
sphere is ruled out.  Systems with declining $\sigma_p(R)$ are
well-matched by theoretical profiles in which the star-forming
galaxies have predominantly radial orbits ($\beta > 0$); many of these
galaxies are probably falling in for the first time. There is
significant evidence for spatial segregation of the spectroscopic
classes regardless of $\sigma_p(R)$.

\end{abstract}

\clearpage

\section{Introduction}

Groups of galaxies have attracted astronomers since the days of
Shapley (1933) and Zwicky (with Humason, 1960). Although the division
between clusters and groups must have seemed natural early
on---comparing the ``rich'' Coma Cluster, for example, with the
``poor'' Stephan's Quintet---today the distinction is hardly
obvious. A host of group catalogs (e.g., Albert, White, \& Morgan
1977; Hickson 1982; Huchra \& Geller 1982; Morgan \& Hartwick 1988;
Ramella, Geller, \& Huchra 1989; Gourgoulhon, Chamaraux, \& Fouque
1992; Ramella, Pisani, \& Geller 1997; Trasarti-Battistoni 1998) make
it clear that ``groups'' are often as well-populated as Abell's (1958)
clusters (to the same apparent magnitude limit), and are sometimes
parts of richer, perhaps bound systems (e.g., Ramella \etal 1994;
Barton, de Carvalho, \& Geller 1998). The potpourri of adjectives with
overlapping meanings still used to describe groups and
clusters---''poor,'' ``rich,'' ``compact,'' and ``loose''--- is a sign
that the relationship among systems of galaxies across the full extent
of their spectrum is not yet well understood.

Much of the literature on ``groups'' deals with an important initial
difficulty in studying poor systems of galaxies: optically selected
catalogs are likely to contain large numbers of apparent systems which
are actually chance superpositions even though the member galaxies
have similar redshifts (e.g., Walke \& Mamon 1989; Hernquist, Katz, \&
Weinberg 1995). Detecting thermal bremsstrahlung x-ray emission from
gas in the group potential increases the probability that galaxies in
an overdense region form a bound system, (e.g., Schwartz, Schwarz, J.,
\& Tucker, 1980; Ponman \etal 1996); however, if the large scale
matter distribution is filamentary, the x-ray emitting gas as well as
the galaxies may be projected onto a deceptively overdense region
(Ostriker, Lubin, \& Hernquist 1995). In conjunction with x-ray
observations, obvious remedies against confusing the chance alignments
with bound systems are to sample the overdense regions more deeply for
additional associated galaxies (e.g., Zabludoff \& Mulchaey 1998), and
to identify additional group members within the neighboring few Mpc of the
overdense region (e.g., Ramella \etal 1995).

There is a rich corpus of theoretical work on the dynamics of
low-velocity dispersion systems. N-body simulations of isolated groups
initialized in virialized states (Bode, Cohn, \& Lugger 1993; Bode
\etal 1994; Athanassoula, Makino, \& Bosma 1997) indicate that
dynamical friction plays an important role in the evolution of groups,
and that the formation of a central, massive galaxy which continually
accretes smaller group members is rapid. In this theoretical scenario
dynamical equilibrium is out of the question because of the large and
roughly constant frequency of galaxy interactions. The cosmological
simulations of Nolthenius, Klypin, \& Primack (1997) similarly show
that only halos with mass $>10^{14} M_\odot$ are guaranteed to be
virialized after a Hubble time; dynamical equilibrium is a poor
description of less massive systems.  However, Bahcall, Gramann, \&
Cen (1994) use their N-body simulations to claim that the velocity
distribution of galaxies in groups is nearly Maxwellian, and
Frederic's (1995b) N-body code provides evidence that certain robust
virial mass estimates applied to groups are accurate and
unbiased. Whatever the truth of the matter, equilibrium models often
form the basis for conclusions about the mass distribution in groups
(e.g., Persic \& Salucci 1992; Dell'Antonio, Geller, \& Fabricant
1995; Pedersen, Yoshi, \& Sommer-Larsen 1997; Zabludoff \& Mulchaey
1998).

Literature which links real groups with theoretical work is less
abundant. There has been some comparison of N-body simulations with
observed group luminosity functions (LFs); Moore, Frenk, \& White
(1993), for example, find good agreement between group LFs and the
results of their N-body simulations. Nolthenius, Klypin, \& Primack
(1994,1997) use other catalog statistics, such as the fraction of
galaxies identified as group members, to constrain the cosmological
parameters from N-body simulations; but Diaferio \etal (1999), whose
simulations distinguish galaxies from dark matter halos, are able to
reproduce the LFs of group catalogs largely independently of the
cosmological parameters. 

For lack of adequate group data, however, direct comparisons of real
systems with theoretical density and velocity dispersion profiles
exist only for rich clusters (e.g. Carlberg \etal 1997a,b). The goal
of this paper is to model the internal dynamics of low-mass systems of
galaxies by providing a catalog suited to the task.

We compare the optical properties of poor groups of galaxies with the
Navarro, Frenk \& White (1997) universal profile, which fits
cosmological simulations of dark matter halos over many orders of
magnitude in halo mass. We also examine the Hernquist (1990) profile,
first derived to fit elliptical galaxies. Throughout the analysis we
divide the system populations according to spectroscopic type by
classifying the individual spectra as either absorption- or
emission-line dominated; in this way we model the kinematics of
galaxies with predominantly older stellar populations separately from
those which have active star formation.

To test the models, it is necessary (1) to have a large enough
($\approx 30$) galaxy membership per system and (2) to sample the
membership well outside the core in order to cover the tails of the
theoretical distributions sufficiently. Previous catalogs of poor
systems fulfill either the first (Pildis \etal 1995; Zabludoff \&
Mulchaey 1998) or the second (Ramella \etal 1995) criterion, but not
both. In \S \ref{sec:data} we describe our assembly and observations
of a catalog of groups, each with membership complete to roughly the
same apparent magnitude ($m_R = 15.4$) within $1.5 h$\m\ Mpc of the
system center; in \S \ref{sec:classify} we discuss an objective method
for spectral classification of the member galaxies.

We identify system members in \S \ref{sec:basics}, and in \S
\ref{sec:vofr} we determine the line-of-sight velocity dispersion
profiles. We examine the behavior of the basic dynamical properties as
a function of spectroscopic type in \S \ref{sec:classtats}, and fit
theoretical models in \S \ref{sec:theory}; in \S \ref{sec:conclusion}
we summarize.

\clearpage

\section{The Data}
\label{sec:data}

We assemble and analyze a sample of galaxy systems, the Deep
Optical Catalog or ``DOC'' hereafter. Two larger surveys are the
sources of the DOC: (1) the Ramella \etal (1999) catalog of galaxy
systems selected optically from the combined CfA--SSRS2 redshifts
surveys (the ``CSOC'' hereafter; see below) and (2) the ROSAT All Sky
Survey---Center for Astrophysics Loose Systems (the ``RASSCALS''
hereafter; Mahdavi \etal 1999), an x-ray emitting subset of the CSOC.

Figure \ref{fig:venn} shows the relationship among the DOC, the CSOC,
and the RASSCALS.  All the systems in the DOC and the RASSCALS also
belong to the CSOC; however the DOC contains more galaxies per system,
and has a high quality optical spectrum for each galaxy.

\subsection{Optical Source Catalog}

The CfA--SSRS2 Optical Catalog (Ramella \etal 1999) is a sample of
galaxy systems drawn objectively from two complete redshift
surveys. It includes systems in a wide variety of dynamical states,
from groups with $\sim 5$ members to the Coma cluster.

The Center For Astrophysics Redshift Survey (Geller \& Huchra 1989;
Huchra \etal 1990; Huchra, Geller, \& Corwin 1995; Falco \etal 1999
CfA) and the smaller Southern Sky Redshift Survey (Da Costa \etal
1994; Da Costa \etal 1998), both complete to $m_B = 15.5$ ($m_R
\approx 14.4$), served as sources for the CSOC. Systems in the CSOC
bear the label of the redshift survey in which they are located; the
names of systems we describe here begin with either NRG (northern CfA)
or SRG (southern CfA). The portion of the CSOC we use covers more than
one sixth of the sky in two separate sections: (1)NRG, $\alpha_{2000}
=$ 8.5 hours to 17 hours and $\delta_{2000} =$ 8.5\degr\ to 43.5\degr,
(2) SRG, $\alpha_{2000} = $ 21.5 hours to 3 hours and $\delta_{2000}
=$ -2\degr\ to 32\degr.

Ramella \etal (1999) use a two-parameter method to construct the
CSOC. Huchra \& Geller (1982) first described this friends-of-friends
algorithm (FOFA) for use with redshift surveys, and Ramella, Pisani,
\& Geller (1997; RPG) revised and applied it to the NRG data. The FOFA
is a three-dimensional algorithm which identifies regions with a
galaxy overdensity $\delta \rho / \rho$ greater than some specified
threshold. A second fiducial parameter, $V_0$, rejects galaxies in the
overdense region which are too far removed in velocity space from
their nearest neighbor. The N-body simulations of Frederic (1995a)
show that the Huchra \& Geller (1982) detection method misses few real
systems, at the cost of including some spurious ones. Ramella \etal
(1999) apply the FOFA to the combined NRG and SRG redshift surveys
with $\delta \rho/\rho = 80.$

The CSOC contains \nlarge\ systems with 3000 km s\m $ \le \vbar \le
12000$ km s\m\ and an average of 9 members galaxies per system. The
low velocity cutoff rejects systems which cover a large area on the
sky, and which may be affected by motions in the Local Group. The
median recession velocity for systems in the CSOC is 7000 km s\m;
therefore the effects of cosmology and evolution are negligible
throughout the sample.

\subsection{The X-Ray Data}

Mahdavi \etal (1997; MBGR) describe a technique to search the ROSAT
All-Sky Survey (Voges 1993; RASS) for low surface brightness x-ray
emission from systems of galaxies. Mahdavi \etal (1999) apply the
method, which uses optical galaxy positions to limit the sky area
searched for emission, to all the groups in the CSOC with 5 or more
members. They use a newly processed version of the survey (RASS II;
Boller \etal 1998), which corrects effects leading to a low detection
rate in the original reduction (RASS I).

The MBGR approach sums x-ray photons only within the union of all
projected $0.4h$\m\ Mpc regions around each system member identified
in the CSOC. This ``position template'' method has an advantage over
traditional detection algorithms: it operates on unbinned, unsmoothed,
raw counts, and as a result can detect much lower surface-brightness
emission. The drawback of the MBGR method is that it can measure only
the total integrated emission within the position template; any
analysis of points sources like active galactic nuclei (AGN) must take
place {\em post facto} with traditional techniques.

Using the position template method, Mahdavi \etal (1999) find that
\nxray\ of the CfA-SSRS2 systems have statistically significant x-ray
emission in the RASS. In each case, they evaluate the probability
$P_{\mathrm point}$ that the x-ray emission is due to a point source
such as an active galactic nucleus (AGN). $P_{\mathrm point}$ results
from a Kolmogorov-Smirnov Test (see \S \ref{sec:KS2D}) comparing the
projected radial profile of the emission with that of the ROSAT PSPC
point spread function. Systems with a high likelihood of having
extended x-ray emission due to bremsstrahlung of hot gas in a
potential ($P_{\mathrm point} \simless 0.03$) make up the catalog of
ROSAT All Sky Survey---Center for Astrophysics Loose Systems
(RASSCALS).

\subsection{The Current Sample of Groups}

The Deep Optical Catalog (Table \ref{tbl:deepcatalog}) consists of
\ndeep\ systems. Of these, \ndeepx\ are selected randomly from the
subset of CSOC groups which (1) have extended ($P_{\mathrm point} \le
0.003$) x-ray emission, (2) have velocity dispersions $\sigma_p < 700$
km s\m, and (3) are observable from the 1.5 meter Tillinghast
Reflector on Mt. Hopkins, Arizona. The remaining systems were selected
randomly from the subset of CSOC groups which fulfill (2) and (3)
above, but either have pointlike or no x-ray emission. The systems
with extended x-ray emission have a higher likelihood of being bound
than the rest, but the presence of this emission is no guarantee that
the groups are virialized.

Because the CSOC systems are identified with objective overdensity and
velocity difference criteria, the DOC is a representative sample of
groups without implicit morphological selection biases. Our sample
includes two Abell clusters, A779 and A1185. Figures
\ref{fig:firstcont}--\ref{fig:lastcont} show the RASSCALS x-ray data
for this sample.

MBGR analyzed the x-ray data for 10 of the DOC systems. However, the
RASSCALS x-ray data differs somewhat from that of MBGR because (1) in
MBGR the fields were not large enough to capture all the x-ray
emission from some of the systems at low redshift, and therefore MBGR
did not detect some of the RASSCALS, and (2) the RASSCALS x-ray data
is from the more current reduction of the ROSAT All Sky Survey (RASS
II). Mahdavi \etal (1999) discuss these issues in greater detail.

We used the high-throughput FAST spectrograph on the Tillinghast
(Fabricant \etal 1998), which, with a 300 line mm\m\ grating and a
$3\as$ slit, has a resolution of $\approx 1.5$ \AA, and a spectral
coverage of 3940 \AA, centered at 5500 \AA. For $z < 0.05$, the
redshift neighborhood of interest for this work, the FAST allows us to
measure the equivalent widths of emission lines which are important
markers of star formation, from [\ion{O}{2}] at 3727 \AA\ to H$\alpha$
at 6563 \AA.

For each system we obtain digitized Palomar Observatory Sky Survey
(POSS) positions for all galaxies with $m_R \simless 15.4$, which lie
within a projected distance of $1.5 h$\m\ Mpc of the system center as
defined in the CSOC. We use the FOCAS (Valdes \etal 1995) program to
separate stars from galaxies, but we also check each field by eye to
make sure FOCAS has not neglected any bright elliptical or edge-on
spiral galaxies. We calibrate the FOCAS magnitudes to the $R$
magnitude system by (1) finding all CSOC galaxies with measured $B$
band magnitudes in each field, (2) applying a standard $B-R = 1.1$ mag
color conversion, the average for elliptical and spirals (Frei \& Gunn
1994) to each CSOC galaxy, and (3) using a linear fit to the
corresponding FOCAS magnitudes to obtain the transformation for all
galaxies in each field. We estimate that the scatter around the $m_R =
15.4$ completeness limit in each system is $\approx 0.3$ mag, and that
from system to system it is $\approx 0.5$ mag.

After obtaining the positions, we measure the spectra of all galaxies
with unknown redshifts, and remeasure the spectra of galaxies with a
known CSOC redshift in the interval $\left| cz - \vbar \right| < 4000$
km s\m, where $\vbar$ is the recession velocity of the system under
study. Therefore the members of all the systems in the DOC, and most
of the foreground and background galaxies, have high-quality FAST
spectra complete to $m_R \approx 15.4$. 

There are a total of \nredtot\ new galaxy spectra in the DOC;
\nredold\ of these are remeasurements of CSOC galaxies. We analyze the
data with the spectroscopy package of the Image Reduction and Analysis
Facility (Valdes 1992; IRAF), and derive redshifts using the RVSAO
program (Kurtz \etal 1992; Kurtz \& Mink 1998), which applies
cross-correlation techniques to log-wavelength binned spectra (Tonry
\& Davis 1979). We use cross correlation for both emission and
absorption spectra.

Table \ref{tbl:data} lists the velocities. A sample page is included
here; the entire table is available in electronic form.

\clearpage

\section{Spectroscopic Classification}
\label{sec:classify}

We use the optical spectrum of each galaxy to decide whether the
galaxy is ``absorption-dominated,'' that is, consists mostly of an
older stellar population (with metallic absorption lines that dominate
the optical spectrum), or whether it is ``emission-dominated'' (with
H$\alpha$, [\ion{O}{2}] and other lines indicative of active star
formation).

To classify each of the new FAST spectra, we first de-redshift it to
the heliocentric rest frame $z = 0$. We fit a cubic spline to
determine the contiuum emission, which we subtract from the
spectrum. Using only the portion between 3600--6900 \AA, binned into
2200 channels of width 1.5 \AA\ each, we fit the linear combination
$C_e T_e + C_a T_a$ to the spectrum, where $C_e$ and $C_a$ are the
dimensionless emission and absorption coefficients determined from the
fit, and $T_e$ and $T_a$ are the template emission and absorption
spectra, respectively.

These templates (Figure \ref{fig:spectmpl}) are the same ones we use
to determine the redshift with RVSAO cross-correlation methods. Kurtz
\& Mink (1998) describe the template construction procedure, which for
$T_e$ consists of averaging the profiles of the emission lines
identified in a large number of emission-dominated galaxy spectra, and
for $T_a$ involves averaging a large number of pure absorption
spectra. The template spectra have no continuum emission. Before
fitting the templates to the FAST data we extract the 3600-6900 \AA\
region. We bin the templates in the same way as we bin the data.

We use the generalized least squares procedure (Press \etal 1992) to
fit the \nfit\ galaxy spectra with 500 km s\m\ $\le c z \le$ 15000 km
s\m. The distribution of $\chi^2/\nu$ for the fits has a mean value of
$1.49$. Figure \ref{fig:coefs} shows shows the distribution of the
parameters $C_e$ and $C_a$. The 68.3\% confidence errors are derived
from the fit covariance matrix.

In Figure \ref{fig:coefs}, the absorption-dominated galaxies cluster
at the lower right corner, where $C_e$ is zero. There is a smooth
transition from this high density region to the emission-dominated
galaxies, which rise towards the upper left corner of the figure. The
emission-dominated population shows no clustering, because absorption
lines from the older stellar population are detectable in almost all the
galaxy spectra; thus most of the galaxies have $C_a \sim 1$.

Separating the two populations of galaxies is not straightforward,
because their distribution does not show a clear break in the
transition zone. Suppose, however, that there is a curve on Figure
\ref{fig:coefs} which divides the populations in some
maximum-likelihood sense. We would like to find this curve as
objectively as possible, keeping in mind that some arbitrariness is
inevitable. Luckily, a few physical considerations constrain the
nature of the division:
\begin{enumerate}
\item The curve must be a function of the absorption coefficient with
positive or zero first derivative. For suppose two galaxies have the
same $C_e$, but different $C_a$; and that the one with the smaller
$C_a$ is classified as absorption-dominated. Then necessarily the
other galaxy must be classified the same way. A similar argument holds
for $C_e$.

\item The curve must not classify galaxies with $C_e \le 0$ as
emission-dominated.

\item The curve must not classify galaxies with ``large enough'' $C_e$
as absorption-dominated.
\end{enumerate}

We separate the two galaxy populations on Figure \ref{fig:coefs} with
a straight line $C_e = a C_a + b$ for simplicity. Condition (1) above
requires $a \ge 0$. Condition (2), given our data, requires, $b > -0.6
a$. Selecting $C_e = \mu_{C_e} + 3 \sigma_{C_e} = 0.3$ as ``large
enough'' (where $\mu_{C_e}$ is the mean and $\sigma_{C_e}$ is the
standard deviation of the \nfit\ $C_e$'s ), condition (3) requires $a
< 1.5$. The permitted region for $a$ and $b$ is now small enough to
allow efficient numerical exploration. 

To find the optimum values of $a$ and $b$, we iterate over the
permitted values, each time classifying galaxies with $C_e \le a C_a +
b$ as absorption-dominated, and the others as emission-dominated. We
exclude two galaxies with anomalous coefficients $C_e \approx C_a
\approx 0$ from this procedure. For each pair of populations we
compute the two-dimensional KS statistic $d_2$ (see \S
\ref{sec:KS2D}), which is a measure of the degree of difference
between two populations; the larger $d_2$, the more likely it is that
the two populations have different distributions.  The maximum value
of $d_2$ therefore gives the maximum-likelihood combination of $a$ and
$b$.

Figure \ref{fig:fitcont} shows a contour plot of $d_2$ in the region
of maximum likelihood. The optimum values are $a = 0.02$ and $b =
0.038$; the fit parameters are stable (to within the uncertainties) as
the sample size increases. Figure \ref{fig:coefs} shows the line which
separates the two populations. About 53\% of the galaxies lie below
the line; these are ``absorption-dominated'' galaxies; the rest are
``emission-dominated.''

\clearpage

\section{Basic Dynamical Properties}
\label{sec:basics}
\label{sec:membership}

Pisani (1993) developed a non-parametric, scale-independent method
which picks out structure in a one-dimensional data set. Without
making use of a histogram, it estimates the probability distribution
function (PDF) underlying the data as a kernel $f_{ka}$ which is a sum
of Gaussians with widths optimized to the local sampling rate:
\begin{eqnarray}
\label{eq:fka}
f_{ka}(v) & \equiv & \frac{1}{N} \sum_i^N K(c z_i, s_i, v) \\
K(c z_i, s_i, v) & \equiv & \frac{1}{\sqrt{2 \pi} s_i} 
\exp\left[-\frac{(v - c z_i)^2}{s_i^2}\right]
\end{eqnarray}
Here $N$ is the number of galaxies in the field, each with a redshift
$z_i$. The kernel is ``adaptive'' because $s_i$
changes for each
data point.

The adaptive kernel $f_{ka}$ is a maximum-likelihood estimator of the
system PDF. Pisani (1993) uses numerical simulations to show that
$f_{ka}$, a function with essentially $N$ degrees of freedom, can
accurately reconstruct even the most asymmetric and non-Gaussian
velocity distributions. The kernel $\fka$ is particularly useful,
because it provides an estimate of the probability that each galaxy in
the field is a member of the system. We can therefore remove
interlopers by discarding all galaxies with a large enough probability
of belonging to the foreground or the background. Furthermore,
whenever $\fka$ detects multiple peaks in the velocity distribution of
a particular system, they are statistically significant substructures
in velocity space.

Figures \ref{fig:hist1}--\ref{fig:hist3} show histograms of the field
of each system alongside $\fka$ for that field. Our procedure for
determining system membership is:
\begin{enumerate}
\item Select all galaxies within $\pm 4000$ km s\m\ of the system
velocity as listed in the CSOC.
\item Compute $\fka$ for these galaxies, as well as the probability
that the $i$th galaxy is an isolated foreground or background object,
$P_{i \in 0} \equiv K(c z_i,s_i,c z_i)/\left[\fka(c z_i) \,
N\right]$. Exclude all galaxies with $P_{i \in 0} > 0.5$ from further
analysis. Note that a galaxy with $P_{i \in 0} < 0.5$ is not
necessarily a system member; it is just ``unisolated.'' While the
$P_{i \in 0} > 0.5$ cutoff is arbitrary, it corresponds well to
traditional rejection algorithms, for example 3--$\sigma$ clipping.
\item Divide $\fka$ into $\nu_{tot}$ peaks by using its minima as
delimiters. The galaxies are therefore also divided into $\nu_{tot}$
bins; the $i$th galaxy belongs to the $\nu$th peak ($i \in \nu$).
\item From $\fka$ compute the probability that the $i$th galaxy
belongs to peak $\nu$:
\begin{equation}
P_{i \in \nu} \equiv \frac{1}{N \fka(c z_i)}
\sum_{j \in \nu, j \neq i} K(c z_j, s_j, c z_i)
\end{equation}
\item Initially define the system membership as the peak $\nu_{max}$
with the most galaxies. However, if neighboring peaks have galaxies
with a non-negligible probability of belonging to the main peak ($P_{
\in \nu_{max}} > 0.003$), include these peaks in the system
membership.
\end{enumerate}
Figures \ref{fig:firstcont}--\ref{fig:lastcont} show sky plots of the
members of each system, along with foreground and background galaxies
and x-ray emission contours. 

After establishing the membership, we compute centers for each system.
If there is a single peak in the x-ray emission and it is associated
with member galaxies, we choose the center of the x-ray peak as the
system center. If, however, the system has multiply-peaked, irregular,
or no x-ray emission, we use the average right ascension and
declination of the optically identified members. In a few cases this
process moves the circle of radius $1.5h$\m\ Mpc away from the CSOC
center, excluding some galaxies with measured spectra from the region
of interest, and therefore from the membership. 

Finally, we compute the system recession velocity $\vbar$ and the
line-of-sight velocity dispersion $\sigma_p$. In \S \ref{sec:robust}
we test various robust methods of deriving these quantities; we find
that the mean velocity $\bar{v}$ and the standard deviation
$\sqrt{\Var(v)}$ are as effective as the robust estimators.

There is a rich qualitative variation in the relationship between the
system membership and nature of the x-ray emission:
\begin{enumerate}

\item Of the \ndeepx\ systems with significant extended x-ray
emission, six (SRGb119, NRGb032, NRGb045, NRGb244, NRGb251, and
SRGb009) are characterized by a single x-ray emitting region which is
relatively round and symmetric. Of these systems, five have either one
or two bright elliptical galaxies at the peak of the x-ray emission:
NRGb032 (two), NRGb244 (one), NRGb251 (one), SRGb009 (two), and
SRGb119 (two). These systems increased their membership to three times
the original CSOC count or better. The system without a bright
elliptical, NRGb045, increased its membership by a factor of 1.6. Thus
there is some indication that the presence of a bright elliptical
galaxy at the x-ray core correlates with richness. The differences in
the population increase are not due to uncertainty in the magnitude
limit.

\item The remaining systems with extended x-ray emission, SRGb062,
NRGb025, NRGs117, NRGb247, NRGs385, and SRGb016, have irregular or
multiple x-ray emission contours; all these systems more than tripled
their original CSOC membership after our deeper sampling.

\item Of the 8 systems without statistically significant extended
x-ray emission, one (NRGb043) more than quintupled its original CSOC
membership; this system is probably bound, and may emit below the RASS
detection threshold. The rest (NRGb007, NRGb004, NRGb057, NRGs127,
NRGs156, NRGb181, and NRGs317) less than doubled the original count.
Because these seven systems lack extended x-ray emission, and did not
change their membership significantly after our survey, they may be
artifacts of the FOFA algorithm. However, the deeper sampling does
reveal a central condensation in both NRGb004 and NRGs317; it is
therefore still not out of the question that these systems might be
real groups which have x-ray emission undetectable in the RASS.

\end{enumerate}

\clearpage

\section{Velocity Dispersion Profiles}
\label{sec:vofr}

\subsection{On the Virial Radius}
\label{sec:r200}

In the literature describing N-body simulations of systems of galaxies
(e.g., Navarro, Frenk, \& White 1997), it is customary to show the
dynamical properties as a function of $r_{200}$, the radius which
contains an overdensity $200 \rho_{\mathrm crit}(z)$ (Navarro, Frenk,
\& White 1997; NFW), where $\rho_{\mathrm crit}(z) = 3 H_0^2(1 + z)^3
/ (8 \pi G) $ is the critical density of an Einstein-de Sitter
universe at a redshift $z$.

Unfortunately, $r_{200}$ is not straightforward to derive
observationally. One method (Carlberg \etal 1997a,b) is to
assume that the virial theorem holds:
\begin{equation}
\label{eq:virial}
M \equiv \alpha \sigma_p^2 r_v G^{-1}.
\end{equation}
Here $r_v$ is the virial radius, $G$ is Newton's gravitational
constant, and $\alpha$ is a constant which depends on the orbit
distribution of the system. Carlberg \etal (1997a,b) use $\alpha = 3$,
appropriate for an isotropic distribution. The next step is to assume
that the mass inside a radius $r$ is proportional to $r$: $M(r)
\propto r$. Then equation (\ref{eq:virial}) with $\alpha = 3$ for a 
system at redshift $z$ implies
\begin{eqnarray}
\frac{\case{4}{3} \pi \rCC^3 \times 200 \rho_{\mathrm crit}(z)}{M_v} &
= & \frac{\rCC}{r_v} \\ 100 \rCC^2 \frac{H_0^2}{G} (1 + z)^3& = &
\frac{M_v}{r_v} \\ \rCC & = & \frac{\sqrt{3} \, \sigma_p (1
+ z)^{-3/2}}{10 \, H_0} \\
& = &  \frac{\sigma_p (1+z)^{-3/2} }{ 577\ {\mathrm km\
s}^{-1}} h^{-1} {\mathrm \ Mpc}.
\label{eq:r200}
\end{eqnarray}
The (usually neglected) $\sim 20\%$ uncertainty in $\sigma_p$
translates into the same error in $\rCC$. This loss of accuracy is
significant considering that practically the only source of error in
the projected radius $R$ is the uncertainty in the determination of
the center, when $R$ is measured in Mpc.

The following are drawbacks in deriving the observational $\rCC$
from the virial theorem:
\begin{enumerate}

\item If the orbits are anisotropic, $\alpha = 3$ is incorrect by up
to a factor of two. 

\item If $M(r)$ is not proportional to $r$, the observational $\rCC$
is inappropriate. NFW's ``universal'' density profile, for example,
predicts $M(r) \propto r^2$ for small radii and $M(r) \propto
\ln{[r/(r_c e)]}$ for large radii.  Carlberg \etal (1997a,b) show that
this profile adequately describes rich clusters of galaxies; below
(see \S \ref{sec:theory}) we find that it also fits our groups well,
whereas an $M(r) \propto r$ profile is ruled out in most cases.

\item Finally, the virial theorem may not apply to some of the systems
in the sample. Even systems with extended x-ray emission are not
guaranteed to be virialized. Substructure generally bloats, and infall
generally compresses the line-of-sight velocity dispersion with
respect to its equilibrium value. For example, the value of $\rCC =
1.2 h$\m\ Mpc for NRGs117 is probably too large, because this system
has three subcondensations (Mahdavi \etal 1996). Similarly, $\rCC =
0.12 h$\m\ Mpc for NRGb045 is unlikely to be a true description of the
overdensity radius, because $\sigma_p = 70$ km s\m, from which it is
derived, is smaller than the internal velocity dispersion of an $L^*$
elliptical galaxy!
\end{enumerate}

\subsection{The Shapes of the Profiles}
\label{sec:profiles}

Figures \ref{fig:hist1}--\ref{fig:hist3} give a qualitative idea of
the velocity dispersion as a function of projected radius. We divide
the data for each group into three bins, containing the galaxies
removed 0--0.5 $h$\m\ Mpc, 0.5--1.0 $h$\m\ Mpc, and 1.0--1.5$h$\m\ Mpc
from the center, respectively, and compute the velocity dispersion for
each of these bins.

Figure \ref{fig:sigofr} shows the velocity dispersion profile
$\sigma_p(R)$ for each system with $N > 10$ members.  We compute
$\sigma_p(R)$ by sorting the members in $R$, and then computing the
velocity dispersion of a moving group of 9 galaxies, from the center
out to a distance of $1.5 h$\m\ Mpc. Every ninth point in the Figure
\ref{fig:sigofr} is statistically independent.  We compute the errors
by performing the bootstrap analysis described in \S
\ref{sec:bootstrap} on each bin.

The behavior of $\sigma_p(R)$ suggests that our sample consists of
systems in a wide variety of dynamical states. The line-of-sight
velocity dispersion profile, for example, rises in certain cases,
falls in other cases, and sometimes just varies irregularly.  The
systems which have the most nearly constant $\sigma_p(R)$---NRGs317,
NRGb057, and NRGb004---are also the systems which have the fewest
members; in these cases we may not be sampling the cores or outer
regions sufficiently to detect a variation. Interestingly, these three
systems also have x-ray emission which has a large probability of
being due entirely to a point source ($P_{\mathrm point} \ge 0.055$),
and therefore have a smaller likelihood of being bound than the
systems with clearly extended x-ray emission.

To examine the variation among the $\sigma_p(R)$ quantitatively, we
perform the KS2D test (see \S \ref{sec:KS2D}) for the \uniquepairs\
unique pairs of systems with $N > 15$ members. Table \ref{tbl:ks}
shows the test results separately for $R$ in units of Mpc and
$r_{200}$. In both cases, for \uniquediss\% of the unique system pairs
in our sample, we can rule out the null hypothesis that $\sigma_p(R)$
is drawn from the same distribution at better than the 99.0\%
confidence level.

The data show, therefore, that our galaxy systems are not, as a rule, in
similar dynamical states, because their velocity dispersion profiles
often differ significantly. If we had followed Zabludoff \& Mulchaey
(1998) or Carlberg \etal (1997), for example, in pooling all our data,
we would have averaged away the distinctive rising, falling, and
irregularly varying individual profiles.

On the one hand, a few of our systems have profiles that qualitatively
match recent N-body simulations of relaxed systems of galaxies as well
as the analytical models we describe below (see \S
\ref{sec:theory}). In SRGb009, SRGb016, NRGb032, SRGb062, and NRGb247,
$\sigma_p(R)$ resembles the velocity dispersion profiles which Crone,
Evrard, \& Richstone (1994) compute for simulated clusters in various
cosmologies. The characteristic rise which all the Crone \etal (1994)
models exhibit within the first $\approx 0.2 \rCC$ is present in
SRGb009 and NRGb247, and $\sigma_p(R)$ declines for all five systems
out to $\approx 2 \rCC$, following the Crone \etal (1994) models. We
will treat these systems, which have consistent distributions
according to the KS2D test, as a special subsample (``Sample-D''). The
x-ray emission in these systems is almost unquestionably extended
($P_{\mathrm point} \le 0.001$), another indication that they
consitute a subsample which lends itself to clean dynamical modeling.

On the other hand, across our sample, $\sigma_p(R)$ does not exhibit
the regularity which it should if all our systems were in dynamical
equilibrium, or if their members had similar families of orbits. Our
results are consistent with recent observational studies of clusters
of galaxies, (e.g. Girardi \etal 1996), which show that even a sample
of systems with $\sigma_p \sim 1000$ km s\m\ exhibit both rising and
falling velocity dispersion profiles within 1.5$h$\m\ Mpc ($\approx 2
h^{-1} \rCC$). Substructure, anisotropies in the true
three-dimensional velocity distribution, and differences in the dark
matter mass profiles, together or individually, may explain the
variations among $\sigma_p(R)$ for different systems of
galaxies.

We suggest that in the systems with declining $\sigma_p(R)$ in the
central region, the galaxies have mainly radial orbits, which bloat
the radial velocity dispersion near the core, and shrink it at the
outer edges. If this is the case, the galaxies could not have made
many orbits; with a typical velocity of $\sim$200--400 km s\m, they
travel only 2$h$\m--4$h$\m\ Mpc in a Hubble time, sufficient to make
just one crossing of the system diameter. It is therefore possible
that a significant fraction of the galaxies are falling towards the
system center for the first time.  In \S \ref{sec:theory} we will
examine this scenario more rigorously, and consider the line-of-sight
velocity dispersion profiles as a function of spectroscopic type.

\clearpage

\section{Spectroscopic Segregation}
\label{sec:classtats}

Elliptical and S0 galaxies in clusters tend to be more centrally
condensed than spiral galaxies, an effect related to the
morphology-density relation in the nearby universe (e.g., Dressler
1980; Postman \& Geller 1984; Lahav \& Saslaw 1992; Whitmore, Gilmore,
\& Jones 1993; Anderson 1996). Some authors (e.g., Carlberg \etal
1997) separate populations by color and refer to ``red'' and ``blue''
galaxies instead.  Sometimes, elliptical or ``red'' galaxies also have
a lower velocity dispersion than the remaining system members (e. g.,
Stein 1996; Colless \& Dunn 1996; Mohr \etal 1996).  Together, these
two effects may imply that many systems perhaps contain virialized
cores of galaxies with older stellar populations, with the
star-forming ``blue'' spirals being more recent, infalling additions
to the system. Our spectral classes correspond only roughly to
morphological or color classifications; however, we seek to establish
whether the stellar populations of galaxies vary with distance from
the system center in a way similar to the morphology or the color.

We compute $\vbar$, $\sigma_p$, and the average distance from the
system center, $\bar{R}$, as a function of spectroscopic
type. Hereafter we use a superscript $a$ with each of these quantities
for absorption-dominated galaxies only, a superscript $m$ for
emission-dominated galaxies only, and no superscript for each system
as a whole. We exclude NRGs127, a system without x-ray emission, from
our tests, because it contains only one emission-dominated galaxy.

We compare the distributions of emission-dominated and
absorption-dominated galaxies with the Student's $t$ Test, which
evaluates the probability that two populations have the same mean, and
the $F$ Test, which evaluates the probability that two populations
have the same variance. When we apply the tests to individual systems,
we use the measured velocities $c z_i$ in km s\m\ and projected
distances $R_i$ in Mpc from the center. Because scaling the data by a
constant does not affect the outcome of either the Student's $t$ or
the $F$ Test, it is only necessary to normalize the velocities and
radii when we compare the absorption-emission populations across
several systems in our sample.

Therefore, when we pool together systems which have similar dynamical
properties, we consider the normalized velocity $v_i/\sigma_p$ in
place of $c z_i$, and measure $R_i$ in units of $\rCC$ as well as
Mpc. We consider several pooled samples: ``ALL'' refers to all the
galaxies which are system members; ``Sample-A'' refers to all the
galaxies except the members of NRGs117 (A1185), a cluster with
significant substructure (Mahdavi \etal 1996) which contains 14\% of
all the member galaxies, and therefore might significantly bias the
test results; ``Sample-D'' refers to the systems with declining
velocity dispersion (\S \ref{sec:profiles}); and ``Sample-I'' refers
to all systems not belonging to Sample-D (and excluding
NRGs117). Table \ref{tbl:classtab} shows the test outcomes; Figure
\ref{fig:distvel} shows the velocity and radius distribution of the
galaxies in each system as a function of spectroscopic type.

The tests unambiguously show that our data recover a ``spectroscopic
segregation'' which resembles the morphology-density relation if we
identify absorption-dominated galaxies with the elliptical or ``red''
members. With better than 99.9\% confidence, $\bar{R}^a$ is different
from $\bar{R}^m$ when we pool all system members together and measure
$R$ in Mpc. On the average, the absorption-dominated galaxies are
$0.14 h$\m\ Mpc closer to the system center than the
emission-dominated galaxies. On an individual basis, nearly all the
systems have $\bar{R}^a - \bar{R}^m < 0$; the two exceptions, NRGb007
and NRGb025, have $\bar{R}^a = \bar{R}^m$ within the uncertainties.
When we measure $R$ in units of $\rCC$, the significance of the effect
for the ALL sample drops to 99.0\%. The physical units (Mpc) preserve
the spectroscopic segregation relation more cleanly; the difference in
physical scale among our systems is therefore smaller than error
introduced into $\rCC$ through the uncertainty in $\sigma_p$; see
equation (\ref{eq:r200}).

The velocity dispersions of the absorption- and emission-dominated
populations, however, show no sign of segregation. NRGs117 is the only
system for which $P(F)$ (the probability that two samples are drawn
from distributions with the same variance) is small. The results of
the all the velocity tests for the combined subsamples are negative.

\clearpage

\section{Comparison with Theory}
\label{sec:theory}

Here we compare the data with spherically symmetric density models
from N-body simulations and the theoretical literature. We have shown
(\S \ref{sec:profiles}) that one particular subsample, Sample-D,
contains groups with statistically consistent velocity dispersion
profiles. These groups also have similar global velocity dispersions
$\sigma_p = 327$--466 km s\m, for which the variation in
$\rCC^{\mathrm Virial}$ is comparable to its uncertainty. While in this
section we focus most of our analysis on Sample-D, we also consider
separately (1) the subsample of high velocity dispersion ($\sigma_p >
350$ km s\m) systems, and (2) the subsample of low velocity dispersion
($\sigma_p < 350$ km s\m) systems.

We consider spherically symmetric density profiles of the form
\begin{equation}
\label{eq:models}
\rho(r) = \frac{3 M_c}{\Xi(\alpha) 4 \pi r_c^3} 
\left[ \frac{r}{r_c} \left(1 + \frac{r}{r_c} \right)^\alpha \right]^{-1}
\end{equation}
where $r_c$ is the core radius, $M_c$ is the mass within the core
radius, and $\Xi(\alpha)$ ensures $\int_0^{r_c} 4 \pi r^2 \rho(r) dr =
M_c$. The model with $\alpha = 3$ is due to Hernquist (1990), who
originally formulated it as a description of elliptical galaxies. The
$\alpha = 2$ model is the universal profile of Navarro, Frenk, \&
White (1996; NFW), who use it as a fitting function to their
simulations of collapsed halos in various cosmologies. The $\alpha =
2$ model is a good fit to the NFW simulations regardless of the
cosmological parameters. The models have normalizations $\Xi(3) = 3/8$
and $\Xi(2) = 3\ln(4/e)/2$, respectively. We also consider the
singular isothermal sphere (SIS), which has $M(r) \propto r$ and
$\rho(r) \propto r^{-2}$.

Carlberg \etal (1997b) also fit these models to their sample of 16
high-luminosity x-ray clusters. Here we compare the models for the
first time to a sample consisting primarily of groups of galaxies.
Our approach differs somewhat from that of Carlberg \etal (1997b):
below we compute the three-dimensional radial velocity dispersion
profile analytically from the Jeans equation, rather than adopting a
fitting function with four free parameters.

For the sake of coherence among the model parameters, our notation for
$\rho(r)$ is somewhat different from that of Hernquist (1990), NFW,
and Carlberg \etal (1997a,b). Because the NFW model has a
logarithmically divergent cumulative mass $M(r)$ as $r \rightarrow
\infty$, NFW define the total halo mass as $M_{200} = 200
\rho_{\mathrm crit} (4 \pi / 3) r_{200}^3$, where $\rCC = c r_c$ is
the radius which contains 200 times the critical density of the
universe, and $c$ is the NFW halo ``concentration.'' NFW find that the
concentration varies inversely with the mass of the halo. We instead
use the mass inside $r_c$, which is more simply defined for both the
NFW and the Hernquist (1990) models, for the profile normalization.

\subsection{Surface Number Density Profiles}

To fit the models to the optical data, we assume that galaxies trace
mass. Then the three-dimensional galaxy number density is
\begin{equation}
\label{eq:nofr}
n(r) = \frac{N_c}{M_c} \rho(r),
\end{equation}
where $N_c$ is the number of galaxies within the sphere of radius
$r_c$.  
Our best fit $N_c$ has a dependence on the $m_R \approx 15.4$
magnitude limit; the best fit $r_c$ should be independent of this
limit if the mass-to-light ratio is constant throughout each
system. The quantity $N_c$ is a superior fitting parameter to the core
density $n_c \equiv 3 N_c / (4 \pi r_c^3)$, because fitting for $n_c$
and $r_c$, as Carlberg \etal (1997a,b) effectively do, results in an
unacceptably large correlation between the parameters. In our
experience, including the directly observable parameters as explicitly
as possible in the models to be fit results in a smaller correlation
among the parameters.

The projection of equation (\ref{eq:nofr}) for $\alpha = 3$ is
the surface number density profile,
\begin{equation}
\Sigma_3(\Rt) = \frac{2 N_c}{\pi r_c^2 \left(\Rt^2 - 1\right)^2}
\left[ \left(2 + \Rt^2 \right) X(\Rt) - 3 \right],
\end{equation}
and for $\alpha = 2$ it is
\begin{equation}
\Sigma_2(\Rt) = \frac{N_c}{\pi \ln(4/e) r_c^2 \left(\Rt^2 - 1 \right)}
\left[1 - X(\Rt) \right],
\end{equation}
where $\Rt = R/r_c$ is the projected radius in units of the core
radius, and
\begin{equation}
X(\Rt) = \frac{\sec^{-1} \Rt}{\sqrt{\Rt^2 - 1}}.
\end{equation}
Note that $X(\Rt)$ is always real; for computation we use ${\mathrm
sech}^{-1} \Rt = i \sec^{-1} \Rt$ when $\Rt < 1$. Also note that
$\Sigma_3(1) = 8 N_c / (15 \pi r_c^2)$ and $\Sigma_2(1) = N_c / [3
\ln(4/e) \pi r_c^2]$. For the singular isothermal sphere, the surface
density has the simple behavior $\Sigma_{\mathrm SIS} \propto R^{-1}$.

To fit models, we use 20 bins, with an approximately equal number of
galaxies $n_{\mathrm bin}$ per bin per fit, so that the fractional
Poisson error $(\sqrt{n_{\mathrm bin}} \ln 10)^{-1}$ in each bin is
roughly constant. However, if this fractional error is greater than
15\%, we decrease the number of bins and increase $n_{\mathrm bin}$
until the fractional error is smaller than 15\%. We conduct two sets
of fits: (a) measuring $R$ in units of $\rCC^{\mathrm Virial}$, given
by equation (\ref{eq:r200}) (in which case $r_c = 1/c$ is
dimensionless and equal to the inverse of the concentration); and (b)
measuring $R$ in units of Mpc. The $\chi^2$ function is well-behaved
and has, in each case, a unique minimum; we compute it over a fine
grid around the minimum, and assume that the large-count Gaussian
limit of the Poisson distribution describes our errors when we show
the 68.3\% and 95.4\% confidence contours. Figures
\ref{fig:firstfit}--\ref{fig:lastfit} show the results of the fits for
Sample-D, which we carry out in the $\log R-\log \Sigma$ plane. We
list the parameters for fits to other subsamples in Table
\ref{tbl:fits}.

The following are the outstanding properties of the fits:
\begin{enumerate}

\item Both the Hernquist (1990) and NFW models provide good
descriptions of the data. The isothermal sphere with $M(r) \propto r$
is usually the worst fit, and is often ruled out.

\item The fits are of much higher quality when we measure $R$ in units
of Mpc than when we use $\rCC^{\mathrm Virial}$; the $\chi^2 / \nu$ is
always smaller. We conclude that the uncertainties discussed above (\S
\ref{sec:r200}) make a large contribution to the $\rCC^{\mathrm
Virial}$, and this quantity is therefore a poor estimator of the true
$\rCC$ for poor systems of galaxies. Classifying systems by the
behavior of $\sigma_p(R)$ (\S \ref{sec:profiles}) is probably a more
effective way to identify dynamically similar groups.

\item The NFW concentration parameter $c$ for Sample-D ($\sigma_p = $
327-466 km s\m) is 4.3-9.1 (one-dimensional 95.4\% confidence
interval); the corresponding value for rich clusters, as derived by
Carlberg \etal (1997a), is 2.3--7.7. In their simulations NFW find 
that low-mass halos should have larger concentrations that high-mass
ones; our results for Sample-D, taken together with those of Carlberg
\etal (1997a,b), are consistent with that prediction.

\item The best-fit core radii $r_c$ for the high- and low-$\sigma_p$
systems would seem to contradict the NFW picture.  If the
low-$\sigma_p$ systems are truly less massive than their
high-$\sigma_p$ counterparts, then NFW predict $c_l > c_h$ as well as
$r_{200,l} < r_{200,c}$. Substituting the definition of the
concentration, we have $r_{c,l} < r_{c,h}$, which is not favored by
the combined fit parameters listed in Table (\ref{tbl:fits}). We are
nevertheless still faced with the fact that the NFW profile fits both
subsamples well. Perhaps the internal inconsistency of the high- and
low-$\sigma_p$ subsamples causes $r_c$ to behave this way. In \S
\ref{sec:profiles} we showed that while Sample-D consists of systems
with well-behaved velocity dispersion profiles, most other systems
have profiles which are inconsistent. Therefore the conclusions drawn
from Sample-D are more robust than those drawn from the high- and
low-$\sigma_p$ subsamples.

\item The fits support the results of our statistical tests in \S
\ref{sec:classtats}. For all our subsamples, the best-fit $r_c$ is
smaller for the absorption- than for the emission-dominated galaxies;
the two-dimensional 68.4\% confidence contours never overlap. 

\end{enumerate}

\subsection{Velocity Dispersion Profiles}

The three-dimensional radial velocity dispersion $\vrthree$ for a
spherical, nonrotating system is given by the Jeans equation (Binney
\& Tremaine 1987):
\begin{equation}
\frac{1}{\rho} \frac{d}{dr} \left( \rho \vrthree \right)
\label{eq:jeans}
+ 2 \beta \frac{\vrthree}{r} = - \frac{G M(r)}{r^2}; \\
\end{equation}
where $\rt = r/r_c$ and $\beta = 1 - \vtthree/\vrthree$ is the
velocity anisotropy parameter. We solve the Jeans equation for the
Hernquist (1990) profile, where
\begin{equation}
M(r) = \frac{4 M_c \rt^2}{(\rt + 1)^2}.
\end{equation}
The Hernquist (1990) density profile yields a much simpler expression
for $\vrthree$ than does NFW model; because both models fit our
spatial data well, we only consider the former. In solving the
equation we use constant $\beta$, rather than assuming some functional
form $\beta(r)$ from N-body simulations as Carlberg \etal (1997b)
do. Because we are not dealing with rich clusters in this work, we
would like freedom from the assumptions that the $\beta$ profiles from
simulations of clusters necessitate, e.g., that the velocity
dispersion is isotropic near the core ($\beta = 0$) and that $\beta$
achieves a maximum somewhere beyond the virial radius (Cole \& Lacey
1996; Diaferio 1999). We therefore seek to establish the average value
of $\beta$ for our systems.

For constant $\beta$ and the Hernquist (1990) profile, equation
(\ref{eq:jeans}) has the solution, 
\begin{equation}
\vrthree = -\frac{4 G M_c}{r_c} \frac{\rtpone^3}{\rt^{2 \beta - 1}}
\int \frac{\rt^{2 \beta - 1}}{\rtpone^5} \, d\rt.
\label{eq:integral}
\end{equation}
While there is no analytic form for the above integral when $\beta$ is
a real number, it is possible to represent the solution as the
following series (see \S \ref{sec:integral}),
\begin{eqnarray}
\label{eq:result1}
\vrthree & = & \frac{G M_c}{r_c} \vrbar(\rt)^2, \\
\vrbar(\rt)^2 & = & \frac{4 \rt}{\left( 5 - 2 \beta \right)
\rtpone^2} \sum_{i = 0}^{\infty} C_i \rtpone^{-i},\\
\label{eq:result2}
C_i & = & \prod_{j = 1}^{i} \frac{5 + j - 1}{5 + j - 2 \beta},
\end{eqnarray}
where $C_0 \equiv 1$. Note that we have set the integration constant
equal to zero to supress divergent solutions as $\rt \rightarrow
\infty$. Equations \ref{eq:result1}--\ref{eq:result2} imply $\lim_{r
\rightarrow 0} \vrthree = \infty$ when $\beta > 0.5$, $\lim_{r
\rightarrow 0} \vrthree = G M_c / r_c$ when $\beta = 0.5$, and
$\lim_{r \rightarrow 0} \vrthree = 0$ when $\beta < 0.5$. The series
always converges for $r > 0$, because $\beta \le 1$; it converges
rapidly when either $\beta < 0.5$ or $\rt > 1$, and slowly when both
$\beta \ge 0.5$ and $\rt < 1$. In computing the series we use double
precision numbers to minimize roundoff error, and add terms until we
achieve a fractional accuracy of $10^{-6}$. For the worst convergence
case ($r \approx 0$, $\beta = 1$) we check the above series
formula with the analytic solution
\begin{equation}
\vrthree = \frac{G M_c}{r_c} \frac{1 + 4 \rt}{3 \rt \rtpone}.
\end{equation}
For this worst-case scenario, the series and the analytic solution
produce identical results to 3 decimal places for $\rt = 0.001$ (which
in practice never enters into consideration), and to 5 decimal places
when $\rt \ge 0.01$.

The projected velocity dispersion profile is (Binney \& Tremaine 1987)
\begin{equation}
\sigma_p^2(R) = \frac{2 N_c}{M_c \Sigma(R)} \int_R^\infty
\left( 1 - \beta \frac{R^2}{r^2} \right)
\frac{\rho \vrthree r dr}{\sqrt{r^2 - R^2}}.
\end{equation}
which we evaluate numerically by reformulating as
\begin{equation}
\sigma_p^2(\Rt) = \frac{4 G M_c N_c}{\pi r_c^3 \Sigma(\Rt)}
\int_0^{1/\Rt} \left( 1 - \beta \Rt^2 t^2 \right)
\frac{\vrbar(1/t)^2\, t^2 dt}{\left(t + 1\right)^3 \sqrt{1 - \Rt^2 t^2}}.
\end{equation}
We integrate the inverse-square-root singularity by using the extended
midpoint rule techniques described in Press \etal (1992; \S 4.4),
which we modify to work with double-precision numbers. Note that once
we specify $r_c$ from fits to the spatial data, the only free
parameters in the theoretical $\sigma_p(R)$ are $\beta$ and
$M_c$.

The combined projected velocity dispersion profiles for the low- and
high-$\sigma_p$ subsamples are too irregular to be good fits to simple
density models. We fit only the combined profile for the well-behaved
set of systems with declining $\sigma_p(R)$ (Sample-D, described in \S
\ref{sec:classtats}). We fix $r_c$ at the value obtained from the
spatial data, and fit $\beta$ and $G M_c / (r_c \sigma_p^2)$ as our
two free parameters. The total mass of the Hernquist (1990) model
results directly from our fit; $M_{\mathrm tot} = 4 M_c$.

Figures \ref{fig:velofr}-\ref{fig:velcont} show the results of the
fits, which have $\chi^2/\nu < 1$ for both the emission- and
absorption-dominated samples taken either individually or together.
Therefore, a Hernquist (1990) profile, with a core radius taken from
spatial data, fits the velocity data well, suggesting that Sample-D is
well described by a spherically symmetric model with a simple velocity
distribution function.

It is perhaps more remarkable that the systems which have non-falling
$\sigma_p(R)$ are still well fit by the simple Hernquist (1990) and
NFW surface density distributions. Our data suggest that while a
spherically symmetric mass distribution provides a good representation
of all our systems, the orbital families are irregularly populated,
and $\beta(r,\theta,\phi)$ has a complicated behavior; it may not be
represented as a spherically symmetric function even though the mass
density is. This, in turn, is further evidence that our systems, with
the exclusion of Sample-D, may not be in dynamical equilibrium.

Because the best-fit $\beta > 0$ for both absorption- and
emission-dominated members in Sample-D, both sets of galaxies are
probably dominated by radial orbits. The one-dimensional 95.4\%
confidence contour for the emission-dominated galaxies never extends
below $\beta = 0.2$; we can therefore rule out orbital isotropy for
them at that level. On the other hand, the orbits of the
absorption-dominated galaxies are consistent with isotropy at the same
level.

Finally, we compute the masses traced by the galaxies.  For Sample-D
as a whole, the 68.3\% confidence interval on the total mass,
not considering systematic effects, is
\begin{equation}
M_{\mathrm tot}  =  (1.2 \pm 0.2)h^{-1} \times 10^{14}  M_\odot
\left( \frac{\sigma_p}{300 {\mathrm \ km\ s^{-1}}} \right)^2. \\
\end{equation}
The separate fits to the absorption- and emission-dominated galaxies 
yield
\begin{eqnarray}
M_{\mathrm tot}^a & = & (1.0 \pm 0.2)h^{-1} \times 10^{14} M_\odot
\left( \frac{\sigma_p}{300 {\mathrm \ km\ s^{-1}}} \right)^2; \\
M_{\mathrm tot}^m & = & (2.0 \pm 0.5)h^{-1} \times 10^{14} M_\odot
\left( \frac{\sigma_p}{300 {\mathrm \ km\ s^{-1}}} \right)^2; \\
\frac{M_{\mathrm tot}^m}{M_{\mathrm tot}^a} & = & 2.0 \pm 0.6.
\end{eqnarray}
The absorption-dominated galaxies trace a mass that is entirely
consistent with Sample-D as a whole. The emission-dominated members
apparently trace a larger mass. The discrepancy in the mass estimates
may be due to fitting $\sigma_p(R)$ for a combined sample rather than
on a case-by-case basis. It is also likely that equilibrium models do
not provide an entirely appropriate description of the
emission-dominated members. If the star-forming galaxies are on
predominantly radial orbits, and are falling in for the first time as
suggested by our timing arguments (\S \ref{sec:profiles}), they
probably have not reached dynamical equilibrium, and are not as well
described by the Jeans equations as the absorption-dominated members.

\clearpage

\section{Conclusions}
\label{sec:conclusion}

We analyze a sample of 20 galaxy systems drawn from a larger,
objectively selected catalog of systems in the Center for Astrophysics
redshift survey. Each system has a membership complete to $m_R
\approx 15.4$ out to a projected radius $1.5h$\m\ Mpc from its center.

\begin{enumerate}
\item Most of the systems have significantly more members after our
observations than they had in the original source catalog (Ramella
\etal 1997). Of the 12 systems with statistically significant,
extended x-ray emission, 11 more than tripled their original
membership. We conclude that our x-ray emitting systems have a high
likelihood of being real, bound groups or clusters. Of the 8 objects
without significant extended emission, one (NRGb043) quintupled its
membership, and may have x-ray emission below the detection threshold
of the ROSAT All Sky Survey; some of the other systems are unlikely to
be bound configurations. 

\item We use fits to template optical spectra, along with the
two-dimensional KS Test (\S \ref{sec:KS2D}), to classify the galaxies
we observe into absorption- and emission-line dominated
populations. We find evidence of spectroscopic segregation: the
members with active star formation have consistently larger mean
distances from the system center than the members with older
stellar populations.  Most systems have many more absorption- than
emission-dominated galaxies in the projected regions encompassed by
the x-ray emission contours.

\item The line-of-sight velocity dispersion as a function of projected
radius, $\sigma_p(R)$, is a useful tool for testing whether our
systems are in similar dynamical states. A quarter of the sample
exhibits declining velocity dispersion profiles similar to the those
implied by N-body simulations (Crone \etal 1994; Navarro, Frenk, \&
White 1997). The remaining 75\% of the systems exhibit flat, rising
and irregularly varying profiles. The null hypothesis that
$\sigma_p(R)$ is drawn from the same distribution is rejected with
better than 99\% significance for \uniquediss\% of all unique pairs of
systems.

\item Both the Hernquist model (1990) and the Navarro, Frenk, \& White
(1997; NFW) universal profile provide good fits to the spatial data,
assuming that galaxies in our sample trace the system mass uniformly.
The isothermal sphere is usually ruled out.  Our best-behaved
subsample (with intermediate velocity dispersions, $\sigma_p=327$--466
km s\m) yields a NFW concentration $c = $ 4.3--9.1 (95.4\% confidence
interval). This is slightly larger than the range for rich clusters
(Carlberg \etal 1997a,b), 2.3--7.7, and therefore consistent with
NFW's prediction that the concentration should decrease as the mass
increases.

\item We solve the Jeans equation for the Hernquist (1990) profile
using a constant velocity anisotropy parameter $\beta$; we project the
solution numerically, and fit the theoretical $\sigma_p(R)$ to the
best-behaved subsample. The theoretical profile provides a good fit,
and the one-dimensional 95.4\% confidence interval for $\beta$ allows
only predominantly radial orbits ($\beta > 0$) for galaxies with
active star formation. Galaxies with older stellar populations have
orbits which are consistent with isotropy to within the errors. The
star-forming galaxies with predominantly radial orbits, especially
those at a distance of $\approx 1.5h$\m\ Mpc, probably have not made
many crossings of the system diameter, since they move only
$2h$\m--$4h$\m\ Mpc in a Hubble time. Many of them may be falling in
for the first time.

\item Remarkably, the systems with irregular or rising velocity
dispersion profiles have a spatial structure which is well-described
by the Navarro, Frenk, \& White (1997) and the Hernquist (1990)
density profiles. Because these systems have surface number densities
which are well-described by theoretical collapsed-halo profiles, but
have velocity structure which no simple model can fit, they are
probably bound configurations which are not in dynamical equilibrium.
\end{enumerate}

In future articles we plan a detailed spectroscopic analysis of
individual system members to measure star formation rates, find
galaxies which show evidence of truncated star formation, and model
the relationship between the spectroscopic properties and the shape of
the x-ray emitting region in our systems.

Thanks are due to the remote observers on the 60$^{\prime\prime}$
Tillinghast Reflector, Perry Berlind, and Michael Calkins, who
together measured many of the spectra for this work. We remember Jim
Peters, who made more than a fourth of the observations, with
fondness; {\em sic itur ad astra}. Brian McLean graciously provided
positions for the galaxies within $28 \le \delta_{2000} \le 32$.  We
could not have done without Susan Tokarz, who did the preliminary
reductions of the data. We thank Gregory Bothun and Ben Moore for
comments which improved the paper. 

This project is based on photographic data of the National Geographic
Society -- Palomar Observatory Sky Atlas (POSS-I) made by the
California Institute of Technology.  The plates were processed into
the present compressed digital form with their permission. The
Digitized Sky Survey was produced at the Space Telescope Science
Institute under US Government grant NAG W-2166. A. Mahdavi,
M. J. Geller, and M. J. Kurtz acknowledge support from the Smithsonian
Institution; A. Mahdavi also acknowledges a National Science
Foundation Graduate Student Fellowship. M. Ramella acknowledges
support from the Italian Space Agency and from the Italian National
Research Council.

\clearpage

\section*{Appendices}
\appendix
\section{Definitions}
\label{sec:definitions}

We use ``system'' to describe any set of galaxies which an objective
structure-finding method places together as a unit. 
A system is ``bound'' if it is self-gravitating; it is an ``artifact''
if the objective method mistakenly placed physically unrelated
galaxies together. The ``membership'' of a system is the set of all
galaxies that have negative total gravitational energy in the center
of mass frame; all other galaxies assigned to the system are
``interlopers.''

A system with $N$ members, each with a measured redshift $z_i$ and
accompanying Gaussian uncertainty $\Delta z_i$, has a recession
velocity $\vbar$, traditionally estimated by $\bar{v} \equiv \sum_i c
z_i/N$, where $c$ is the speed of light. The $i$th member has a
peculiar line-of-sight velocity $v_i \equiv \left( c z_i - \vbar
\right) / \left(1 + \vbar/c\right)$. The global projected velocity
dispersion of the system is $\sigma_p$, where traditionally
$\sigma_p^2$ is estimated by $\Var(v) \equiv \sum_i \left( v_i - \vbar
\right)^2 / (N - 1)$. The standard deviation is defined as
$\sqrt{\Var(v)}$.  In \S \ref{sec:robust} we consider robust estimators
of $\vbar$ and $\sigma_p$.

We indicate three-dimensional distances from the system center with a
lowercase $r$, while reserving the uppercase $R$ for projected
distance on the sky. In our article the ``field'' of a system is a
circle on the sky with the same center as the system, and with radius
$\theta \equiv 150 $ km s\m$/\vbar$. In other words, the field is
always large enough to include a $R = 1.5 h$\m\ Mpc region around the
system center.

\section{On Bootstraps With Measurement Errors}
\label{sec:bootstrap}

When the probability distribution underlying the data, $D(x)$, is {\em
a\ priori} unknown---as is the case with the velocity distribution of
system members---it is not possible to compute confidence
intervals analytically for some parameter, $a$, derived from the
data. The bootstrap is a numerical method for obtaining these
confidence intervals by using the data set itself as an estimate of
the underlying distribution (see, e.g., Lupton 1993). Classically, for
a data set $x_i$ of length $N$,
\begin{equation}
D(x) \equiv \frac{1}{N} \sum_i^N \delta (x - x_i)
\end{equation}
Here $\delta(x)$ is the Dirac delta function. To compute the
confidence interval on $a$, one usually draws a large number of data
sets of length $N$ from $D(x)$, and by recomputing $a$ each time,
derives its probability distribution. For the case that $D(x)$ is a
sum of delta functions, this procedure is tantamount to selecting
points from the data set with replacement.

Unfortunately, by representing each data point as a delta function,
$D(x)$ neglects the uncertainty accompanying that measurement. The
probability distribution of each individual measurement is not
$\delta(x - x_i)$, but some error distribution $E(x)$ centered on the
measurement. In the case of our redshift measurements, $E(z)$ is a
Gaussian with a standard deviation $\Delta z_i \approx
10^{-4}$. Therefore the correct bootstrap data probability
distribution is
\begin{equation}
D(z) 
\equiv \frac{1}{\sqrt{2 \pi} N} 
\sum_i^N \frac{1}{\Delta z_i}\exp
\left[-\frac{(z - z_i)^2}{\Delta z_i^2}\right]
\end{equation}
Drawing deviates distributed as $D(z)$ is a bit more tricky than
selecting with replacement, as is possible with the delta function
distribution. We use a rejection method, as outlined for example in
Press \etal (1992, \S 7.3), with a constant comparison function over
the range of the data. All the bootstrap confidence intervals for the
parameters we derive therefore take the measurement uncertainties into
account properly.

Note that $D(x)$, $D(z)$ and the adaptive kernel estimator $\fka$,
defined in equation (\ref{eq:fka}), are only superficially
related. The first two are minimal representations of the distribution
of the data, constructed for the purpose of deriving alternate data
sets drawn from a similar parent population. The adaptive kernel, on
the other hand, attempts to overcome the effects of sampling as it
builds its guess at the true shape of the parent distribution.

\section{On the Two-Dimensional KS Test}
\label{sec:KS2D}

The Kolmogorov-Smirnov Test (e.g., Press \etal 1992, \S 14.3)
evaluates the null hypothesis that a pair of data sets are drawn from
the same distribution. In the simple case that each data set is one
dimensional, the correponding KS Test (KS1D) computes $d_1$, the
maximum deviation of the two data cumulative distributions.  The {\em
a priori} distribution of $d_1$ itself is known, and therefore so is
the probability of the null hypothesis. The KS1D test is particularly
useful, because its estimate of the probability is independent of the
original distributions of the data, which are often not known.

We use a version of the Kolmogorov-Smirnov test (KS2D; e.g. Press
\etal 1992, \S 14.7) which is applicable to two-dimensional data
sets. Because the notion of a cumulative distribution is poorly
defined for a two-dimensional data set, the corresponding deviation
statistic, $d_2$, is calculated somewhat differently. Suppose we are
comparing two data sets, $(x_i, y_i)$, $i = 1 \ldots N_1$ and
$(X_j,Y_j)$, $j = 1 \ldots N_2$. For each $(x_i,y_i)$ compute
\begin{eqnarray}
\nonumber
{\mathcal D}_i & \equiv \mathrm{max\ } & \left(
       \left|\sum_{x < x_i,\ y < y_i}\frac{1}{N_1} -
	     \sum_{X < x_i,\ Y < y_i}\frac{1}{N_2} \right|, 
       \left|\sum_{x > x_i,\ y < y_i}\frac{1}{N_1} -
	     \sum_{X > x_i,\ Y < y_i}\frac{1}{N_2} \right|, \right. \\
&&\left.\ \ 
       \left|\sum_{x < x_i,\ y > y_i}\frac{1}{N_1} -
	     \sum_{X < x_i,\ Y > y_i}\frac{1}{N_2} \right|, 
       \left|\sum_{x > x_i,\ y > y_i}\frac{1}{N_1} -
	     \sum_{X > x_i,\ Y > y_i}\frac{1}{N_2} \right| \right).
\end{eqnarray}

Define ${\mathcal D}_j$ similarly, replacing $i$ with $j$, $x_i$ with
$X_j$, and $y_i$ and $Y_j$ in the above
equation. Then
\begin{equation}
d_2 \equiv \left( \frac{N_1 N_2}{N_1 + N_2} \right)^{\frac{1}{2}}
\times \frac{{\mathrm max} ({\mathcal D}_i) + {\mathrm max}
({\mathcal D}_j)}{2}
\end{equation}

To compute $d_2$ we use the computer code which Press \etal (1992)
implement on the basis of theoretical work by Fasano \& Franceschini
(1987). However, we do not use the Press \etal (1992) prescription to
compute the probability of the null hypothesis, because it is only
accurate for $N \simgreat 20$ data points. Rather we use the results
of Fasano \& Franceschini (1987) directly; they fit a third-order
polynomial in three variables (sample size, sample correlation, and
desired significance level) to their Monte Carlo simulations of the
probability distribution of $d_2$. They find that the fractional
uncertainty of the resultant probabilities for the null hypothesis is
$\sim 5\%$.

Note that the output $d_P$ of the Press \etal (1992) computer program
is normalized differently from our notation for $d_2$, which agrees
with that of Fasano \& Franceschini (1987):
\begin{equation}
d_2 = d_P \left( \frac{N_1 N_2}{N_1 + N_2} \right)^{\frac{1}{2}}.
\end{equation}

\section{Comparison of the Robust and Classical Estimators of 
the Velocity Moments}
\label{sec:robust}

The classical estimators of the system recession velocity $\vbar$ and
the line-of-sight velocity dispersion $\sigma_p$ are the mean velocity
$\bar{v}$ and the unbiased standard deviation estimator
$\sqrt{\Var(v)}$, respectively (see \S
\ref{sec:definitions}). However, Beers, Flynn, \& Gebhardt (BFG; 1990)
argue that the sample mean and the sample standard deviation are not,
in general, efficient estimators of the properties of a parent
distribution $p(x)$. For example, if $p(x)$ is not a Gaussian,
$\bar{x}$ is in general a poorer estimator of $\langle x \rangle$, and
$\Var(x)$ is a poorer estimator $\langle \left( x - \langle x \rangle
\right)^2 \rangle$, than certain robust estimators which we compute
alongside the classical estimators.

In Figure \ref{fig:robust} we show $\bar{v}$ alongside the sample
median $M$, and the biweight location estimator,
\begin{eqnarray}
{\mathrm BiMean} & = & M +
\frac{\sum_{\left| \mu_i \right| < 1} 
\left(c z_i - M \right)
\left(1 - \mu_i^2 \right)^2}
{\sum_{\left| \mu_i \right| < 1} \left(1 - \mu_i^2 \right)}; \\
\mu_i & = & \frac{\left(x_i - M \right)}{c \times {\mathrm MAD}}; \\
{\mathrm MAD} & = & {\mathrm median} \left(\left| x_i - M \right|\right).
\end{eqnarray}
Here $c$ is an arbitrary tuning parameter. Similarly, we list the
standard deviation alongside the ``Gapper,'' which uses the weighted
gaps among the data points to compute the scale of the velocity
distribution (see BFG for its definition); we also show the biweight
scale estimator,
\begin{equation}
{\mathrm BiSigma} = n^{1/2}
\frac{\left[\sum_{\left| \mu_i \right| < 1} 
\left(c z_i - M \right) \left(1 - \mu_i^2 \right)^4\right]^{1/2}}
{\left| \sum_{\left| \mu_i \right| < 1} 
\left(1 - \mu_i^2 \right) \left(1 - 5 \mu_i^2 \right) \right|}
\end{equation}

Interestingly, these various estimators differ very little (Figure
\ref{fig:robust}). For all systems, the robust biweight estimates of
the recession velocity and velocity dispersion lie well within the
68.3\% confidence interval of the classical estimators, the mean and
the standard deviation. The median, which is not efficient in the
presence of substructure (BFG), lies outside the confidence interval
of the mean only twice. The even closer agreement of the estimators of
$\sigma_p$, a quantity somewhat more important for kinematic analysis
than $\vbar$, is also noteworthy.

In considering which of the two kinds of estimators, classical or
robust, we should use, we note that the latter make use of a
dimensionless tuning constant $c$, which BFG set to 6.0 for the
biweight location estimator, and to 9.0 for the scale estimator. The
biweights also require the subjective choice of an auxiliary estimator
of scale; BFG use the mean absolute deviation (``MAD''). On the other
hand the classical estimators contain no arbitrary parameters.

We therefore adopt $\vbar = \bar{v}$ and $\sigma_p = \sqrt{\Var(v)}$.
All the quoted errors on the velocity moments are 68.3\%
bias-corrected bootstrap confidence intervals as described in \S
\ref{sec:bootstrap}.

\section{Integration of Constant-$\beta$ Jeans Equation}
\label{sec:integral}

Equation (\ref{eq:integral}) is integrable through the identity,
\begin{equation}
\int x^m (a + b x)^n \, dx = \frac{x^{m+1} (a + b x)^{n}}
{m + n + 1} + \frac{an}{m + n + 1} \, \int x^m (a + b x)^{n-1} \, dx.
\end{equation}

Substituting $a = b = 1$, $x = \rt$, $m = 2 \beta - 1$, and $n = -5$,
and applying the above expression to the integral in equation
\ref{eq:integral}, we obtain
\begin{eqnarray}
\int \frac{\rt^{2 \beta - 1}}{(\rt + 1)^5} \, d\rt & =  &
\frac{1}{2 \beta - 5} \left\{ \frac{\rt^{2 \beta}}{(\rt + 1)^5}
 - \frac{5}{2 \beta - 6} \left[ \frac{\rt^{2 \beta}}{(\rt + 1)^6}
 - \frac{6}{2 \beta - 7} \left( \frac{\rt^{2 \beta}}{(\rt + 1)^7} 
- \ldots \right) \right] \right\}, \\
&=& -\frac{\rt^{2 \beta}}{(\rt + 1)^5} \frac{1}{5 - 2 \beta}
\left\{ 1
 + \frac{5}{6 - 2 \beta} \left[ \frac{1}{(\rt + 1)}
 + \frac{6}{7 - 2 \beta} \left( \frac{1}{(\rt + 1)^2} 
 + \ldots \right) \right] \right\}.
\end{eqnarray}
From the above relation, equations \ref{eq:result1}--\ref{eq:result2}
follow easily.

\parindent 0in
\clearpage
\renewcommand{\baselinestretch}{1.0}
\section*{References}
\addcontentsline{toc}{section}{References}

Abell, G. O. 1958, ApJS, 3, 211

Albert, C. E., White, R. A., \& Morgan, W. W. 1977, ApJ, 211, 309

Anderson, S. 1996, A\&A, 314, 763

Antonuccio-Delogu, V., \& Colafrancesco, S. 1994, ApJ, 427, 72

Athanassoula, E., Makino, J., \& Bosma, A. 1997, MNRAS, 286, 825

Barton, E., de Carvalho, R. R., \& Geller, M. J. 1998, AJ, in press

Beers, T. C., Flynn, K., \& Gebhardt, K. 1990, AJ, 100, 32

Binney, J., and Tremaine, S. 1987, Galactic Dynamics (Princeton:
Princeton University Press)

Bode, P. W., Cohn, H. N., \& Lugger, P. M. 1993, ApJ, 416, 17

Bode, P. W., Berrington, R. C., Cohn, H. N., \& Lugger, P. M. 1994,
ApJ, 433, 479

Boller, T., Bertoldi, F., Dennefeld, M., \& Voges, W. 1998, A\&AS,
129, 87

Carlberg, R. G., Yee, H. K. C., Ellingson, E., Morris, S. L., Abraham,
R., Gravel, P., Pritchet, C. J., Smecker-Hane, T., Hartwick, F. D. A.,
Hessler, J. E., Hutchings, J. B., \& Oke, J. B. 1997a, ApJ, 476, L7

Carlberg, R. G., Yee, H. K. C., Ellingson, E., Morris, S. L., Abraham, 
R., Gravel, P., Pritchet, C. J., Smecker-Hane, T., Hartwick, F. D. A., 
Hessler, J. E., Hutchings, J. B., \& Oke, J. B. 1997b, ApJ, 495, L13

Cole, S., \& Lacey, C. G. 1996, MNRAS, 281, 716

Colless, M., \& Dunn, A. M. 1996, ApJ, 458, 435

Crone, M. M., Evrard, A. E., \& Richstone, D. O. 1994, ApJ, 434, 402

Da Costa, L. N., Geller, M. J., Pellegrini, P. S., Latham,
D. W., Fairall, A. P., Marzke, R. O., Willmer, C. N. A., Huchra,
J. P., Calderon, J. H., Ramella, M., and Kurtz, M. J. 1994, ApJ, 424,
L1

Da Costa, L. N., Willmer, C. N. A., Pellegrini, P. S., Chaves, O. L., 
Maia, M. A. G., Geller, M. J., Latham, D. W., Kurtz, M. J., Huchra, 
J. P., Ramella, M., Fairall, A. P., Smith, C., \& Lipari, S. 1998, AJ,
116, 1

Dell' Antonio, I. P., Geller, M. J., \& Fabricant, D. G. 1994, AJ,
107, 427

Diaferio, A., Ramella, M., Geller, M. J., Ferrari, A. 1993, AJ, 105, 2035

Diaferio, A., Geller, M. J., \& Ramella, M. 1994, AJ, 107, 868

Diaferio, A., Kauffmann G., Colberg, J. M., and White, S. D. M. 1999,
submitted to MNRAS

Diaferio, A. 1999, submitted to ApJ

Dressler, A. 1980, ApJ, 236, 351

Dressler, A., \& Shectman, S. A. 1998, AJ, 95, 985

Fabricant, D., Cheimets, P., Caldwell, N., and Geary, J. 1998, PASP,
110, 79

Falco, E. E., Kurtz, M. J., Geller, M. J., Huchra, J. P.,
Peters, J., Berlind, P., Mink, D. J., Tokarz, S. P., and Elwell,
B. 1999, PASP, in press

Fasano, G., \& Franceschini, A. 1987, MNRAS, 225, 155

Frederic, J. 1995a, ApJS, 97, 259

Frederic, J. 1995b, ApJS, 97, 275

Frei, Z., \& Gunn, J. E., 1994, AJ, 111, 174

Geller, M. J., \& Huchra, J. P. 1989, Science, 246, 897

Girardi, M., Fadda, D., Guiricin, G., Mardirossian, F., Mezzetti,
M., \& Biviano, A. 1996, ApJ, 457, 61

Gourgoulhon, E., Chamaraux, P., \& Fouque, P. 1992, A\&A, 255, 69

Hernquist, L. 1990, ApJ, 356, 359

Hernquist, L., Katz, N., \& Weinberg, D. H. 1995, ApJ, 442, 57

Hickson, P. 1982, ApJ, 255, 382

Huchra, J. P., \& Geller, M. J. 1982, ApJ, 257, 423

Huchra, J. P., de Lapparent, V., Geller, M. J., \& Corwin, Jr., H. G.
1990, ApJS, 72, 433

Huchra, J. P., Geller, M. J., \& Corwin, Jr., H. G. 1995, ApJS, 70, 687

Kurtz, M. J., Mink, D. J., Wyatt, W. F., Fabricant, D. G., Torres, G.,
Kriss, G. A., and Tonry, J. L. 1992, in Astronomical Data Analysis
Software and Systems I, A. S. P. conference series, Vol. 25, Worrall,
D. M., Biemesderfer, C., and Barnes, J., eds., p. 432

Kurtz, M. J., \& Mink, D. J. 1998, PASP, 110, 934

Lahav, O., \& Saslaw, W. C. 1992, ApJ, 396, 430

Lupton, R. 1993, Statistics in Theory and Practice (Princeton:
Princeton University Press)

Mahdavi, A., Geller, M. J., Fabricant, D. G., Kurtz, M. J., Postman,
M., and McLean, B. 1996, AJ, 111, 64

Mahdavi, A., B\"ohringer, H., Geller, M. J., \& Ramella, M. 1997,
ApJ, 483, 68

Mahdavi, A., B\"ohringer, H., Geller, M. J., \& Ramella, M. 1999, in
preparation

Marzke, R. O., Huchra, J. P., \& Geller, M. J. 1993, ApJ, 428, 43

Menci, N., \& Fusco-Femiano, R. 1996, ApJ, 472, 46

Mohr, J. J., Geller, M. J., Fabricant, D. G., Wegner, G.,
Thorstensen, J., \& Richstone, D. O. 1996, ApJ, 470, 724

Moore, B., Frenk, C. S., \& White S. D. M. 1993, MNRAS, 261, 827

Kurtz, M. J., \& Mink, D. J. 1998, PASP, in press

Morgan, C. G., \& Hartwick, F. D. A. 1988, ApJ, 328, 381

Mulchaey, J. S., \& Zabludoff, A. I. 1998, ApJ, 496, 73

Navarro, J. F., Frenk. C. S., \& White, S. D. M. 1997, ApJ, 490, 493

Nolthenius, R., Klypin, A. A., \& Primack, J. R. 1994, ApJ, 422, L45

Nolthenius, R., Klypin, A. A., \& Primack, J. R. 1997, ApJ, 480, 43

Ostriker, J. P., Lubin, L. M., \& Hernquist, L. 1995, ApJ, 444, L61

Pedersen, K., Yoshi, Y., \& Sommer-Larsen, J. 1997, ApJ, 485, L17

Persic, M., \& Salucci, P. 1992, MNRAS, 258, 14

Pisani, A. 1993, MNRAS, 265, 706

Pisani, A. 1996, MNRAS, 278, 697

Ponman, T. J., Bourner, P. D. J., Ebeling, H., \& B\"ohringer, H.
1996, MNRAS, 283, 690

Postman, M., Geller, M. J. 1984, ApJ, 281, 95

Press, W. H., Teukolsky, S. A., Vetterling, W. T., and Flannery,
B. P. 1992, Numerical Recipies in C, Second Edition (Cambridge,
England: Cambridge University Press)

Ramella, M., Geller, M. J., \& Huchra, J. P. 1989, ApJ, 344, 57

Ramella, M., Geller, M. J., Huchra, J. P., \& Thorstensen, H. R.,
1995, AJ, 109, 1458 (RGHT)

Ramella, M., Pisani, A., \& Geller, M. J. 1997, AJ, 113, 483 (RPG)

Ramella, M., \etal 1999, in preparation

Shapley, H. 1933, Proc. Nat. Acad. Sci. 19, 591

Schechter, P. 1976, ApJ, 203, 297

Schwartz, D. A., Schwarz, J., and Tucker, W. 1980, ApJ, 238, L59

Stein, P. 1997, A\&A, 317, 670

Tonry, J. L., \& Davis, M. 1979, AJ, 84, 1511

Trasarti-Battistoni, R. 1998, A\&AS, 130, 341

Valdes, F. 1992, in Astronomical Data Analysis Software and Systems,
A. S. P. conference series, Vol. 25, Worrall, D. M., Biemesderfer, C.,
and Barnes, J., eds., p. 417

Valdes, F. G., Campusano, L. E., Velasquez, J. D., \& Stetson, P. B.
1995, PASP, 107, 1119

Vio, R., Fasano, G., Lazzarin, M., and Lessi, O. 1994, A\&A, 289, 640

Voges, W. 1993, Advances in Space Research, 13, (12)391

Walke, D. G., \& Mamon, G. A. 1989, A\&A, 225, 291

Whitmore, B. C., Gilmore, D. M., \& Jones, C. 1993, ApJ, 407, 489

Zabludoff, A. I., \& Mulchaey, J. S. 1998, ApJ, 496, 39

Zwicky, F., \& Humason, M. L. 1960, ApJ, 132, 627

\clearpage

\plotone{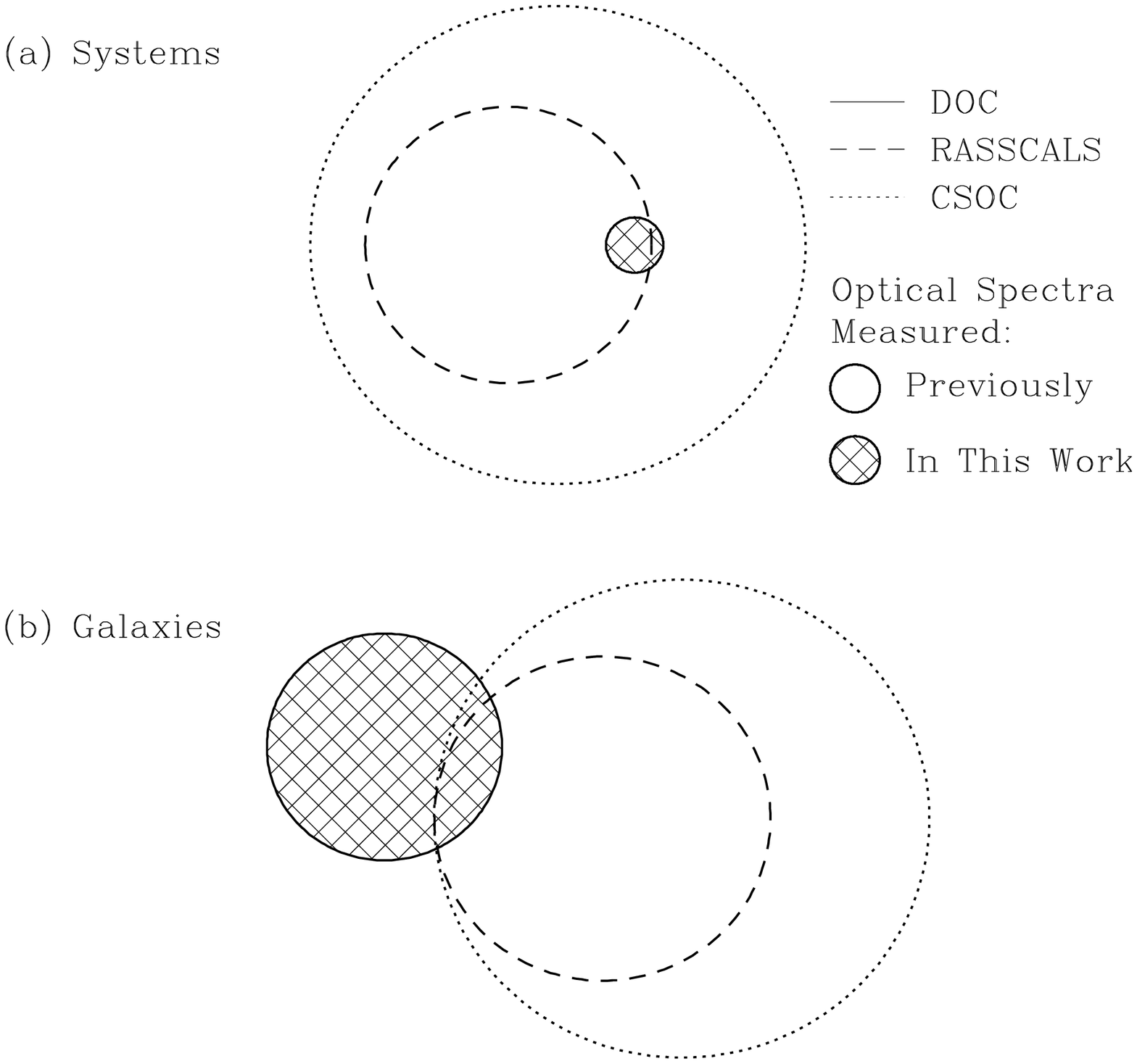}
\figcaption[venn.eps]{Venn diagrams showing the relationship among the
Deep Optical Catalog (This paper), the RASSCALS (Mahdavi \etal 1999),
and the CfA-SSRS2 Optical Catalog (Ramella \etal 1999). In (a), the
circles represent the set of all galaxy systems in each catalog. The
area of each circle is proportional to the number of systems it
contains; the areas of the intersections accurately represent the
degree of overlap among the catalogs. In (b), the circles represent
the set of all galaxies classified as system members in each
catalog. The area of each circle is proportional to the total number
of galaxies classified as members, and the areas of the intersections
accurately reflect the number of galaxies the catalogs share.
\label{fig:venn}}

\epsscale{1.0}
\plottwo{SRGb062.epsi}{SRGb119.epsi}

\ 

\plottwo{NRGb004.epsi}{NRGb007.epsi} 

\figcaption{Galaxy positions and x-ray contours. The x-ray emission
contours begin at $1.33$ standard deviations above the level of the
background, and increase by a factor of 1.5; dotted contours lie less
than 3.0 standard deviations above the background. Galaxies with $c z
> $ 15000 km s\m\ are simply shown as crosses; they are never system
members. Galaxies with 500 km s\m\ $ \le c z \le $ 15000 km s\m\ are
marked by circles if their spectra are absorption dominated, or by
triangles if the spectra are emission-dominated. Each triangle or
circle is filled-in if the galaxy is a system member, empty if it is
not.
\label{fig:firstcont}}

\plottwo{NRGb025.epsi}{NRGb032.epsi}

\ 

\plottwo{NRGb043.epsi}{NRGb045.epsi}

\figcaption[NRGb025.epsi,NRGb032.epsi,NRGb043.epsi,NRGb045.epsi]
{See caption for Figure \protect\ref{fig:firstcont}.
\label{fig:secondcont}}

\plottwo{NRGb057.epsi}{NRGs117.epsi}

\

\plottwo{NRGs127.epsi}{NRGs156.epsi}

\figcaption[NRGb057.epsi,NRGs117.epsi,NRGs127.epsi,NRGs156.epsi]
{See caption for Figure \protect\ref{fig:firstcont}.
\label{fig:NRGs156}}

\plottwo{NRGb181.epsi}{NRGb244.epsi}

\

\plottwo{NRGb247.epsi}{NRGb251.epsi}

\figcaption[NRGb181.epsi,NRGb244.epsi,NRGb247.epsi,NRGb251.epsi]
{See caption for Figure \protect\ref{fig:firstcont}.}

\plottwo{NRGs317.epsi}{NRGs385.epsi}

\

\plottwo{SRGb009.epsi}{SRGb016.epsi}

\figcaption[NRGs317.epsi,NRGs385.epsi,SRGb009.epsi,SRGb016.epsi]
{See caption for Figure \protect\ref{fig:firstcont}.	
\label{fig:lastcont} \label{fig:NRGs385}}

\plotone{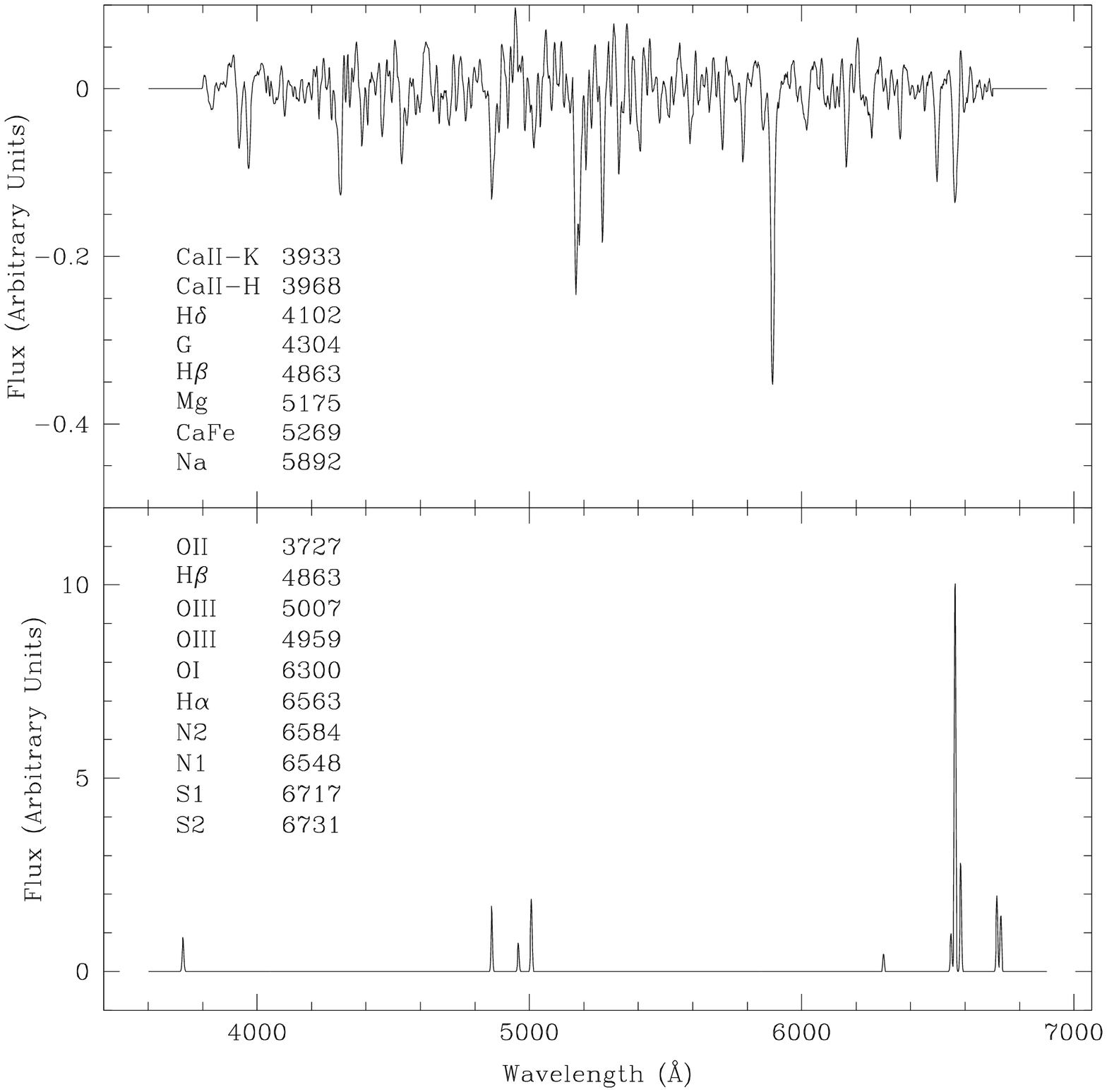}
\figcaption[spectmpl.eps]{The absorption and emission spectral
templates, with a list of the lines present.
\label{fig:spectmpl}}

\epsscale{1.0}
\plotone{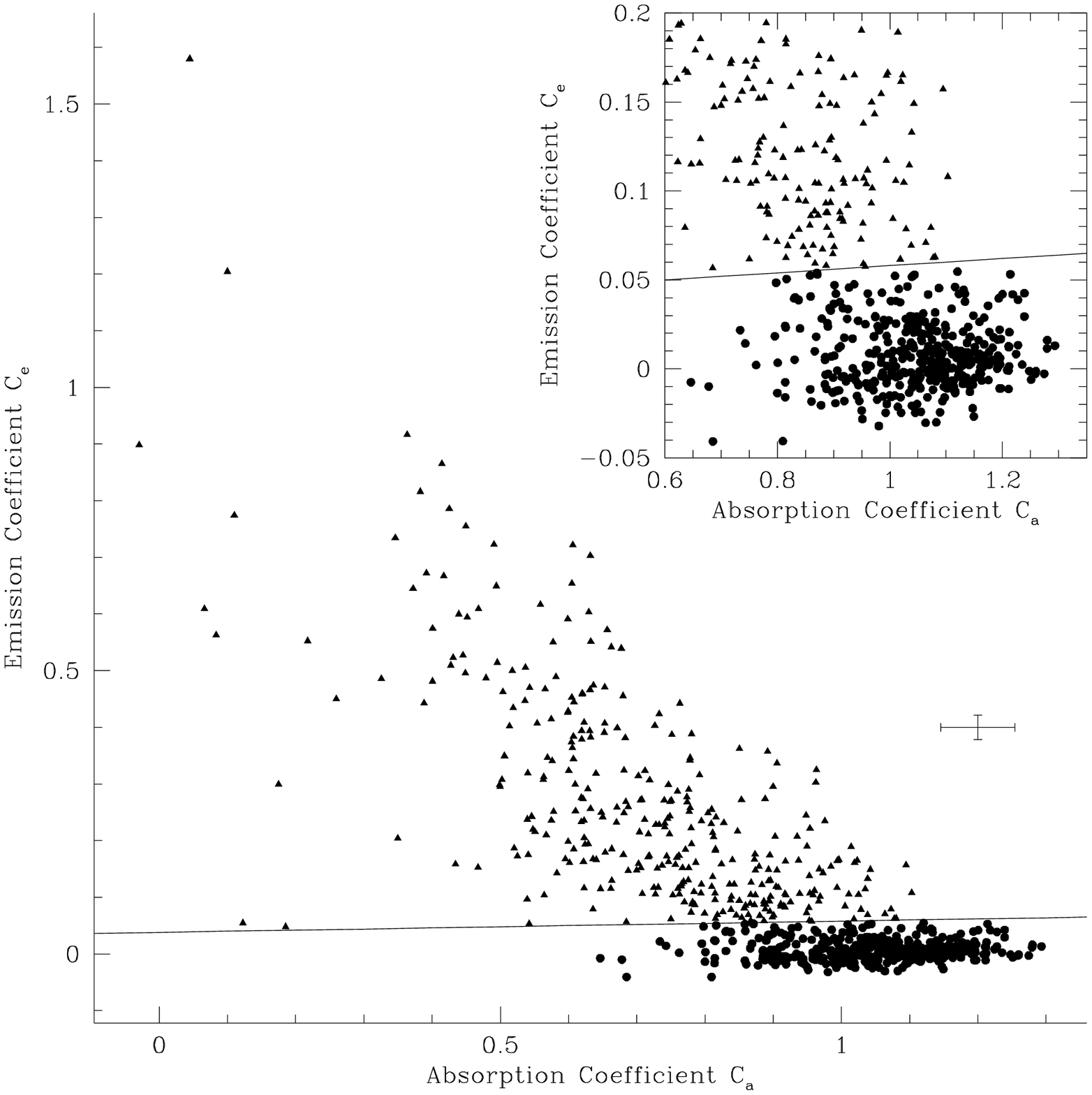}
\figcaption[coefs.eps]{$C_a$ vs. $C_e$ for the \protect\nfit\ galaxies
with 500 km s\m\ $ \le c z \le$ 15000 km s\m. Triangles and circles represent
emission-dominated and absorption-dominated galaxies, respectively.
The errorbars show the typical uncertainty in the classification;
the straight line is the maximum-likelihood dividing line described
in the text. The top right box is a magnification of the
absorption-dominated region.
\label{fig:coefs}}

\plotone{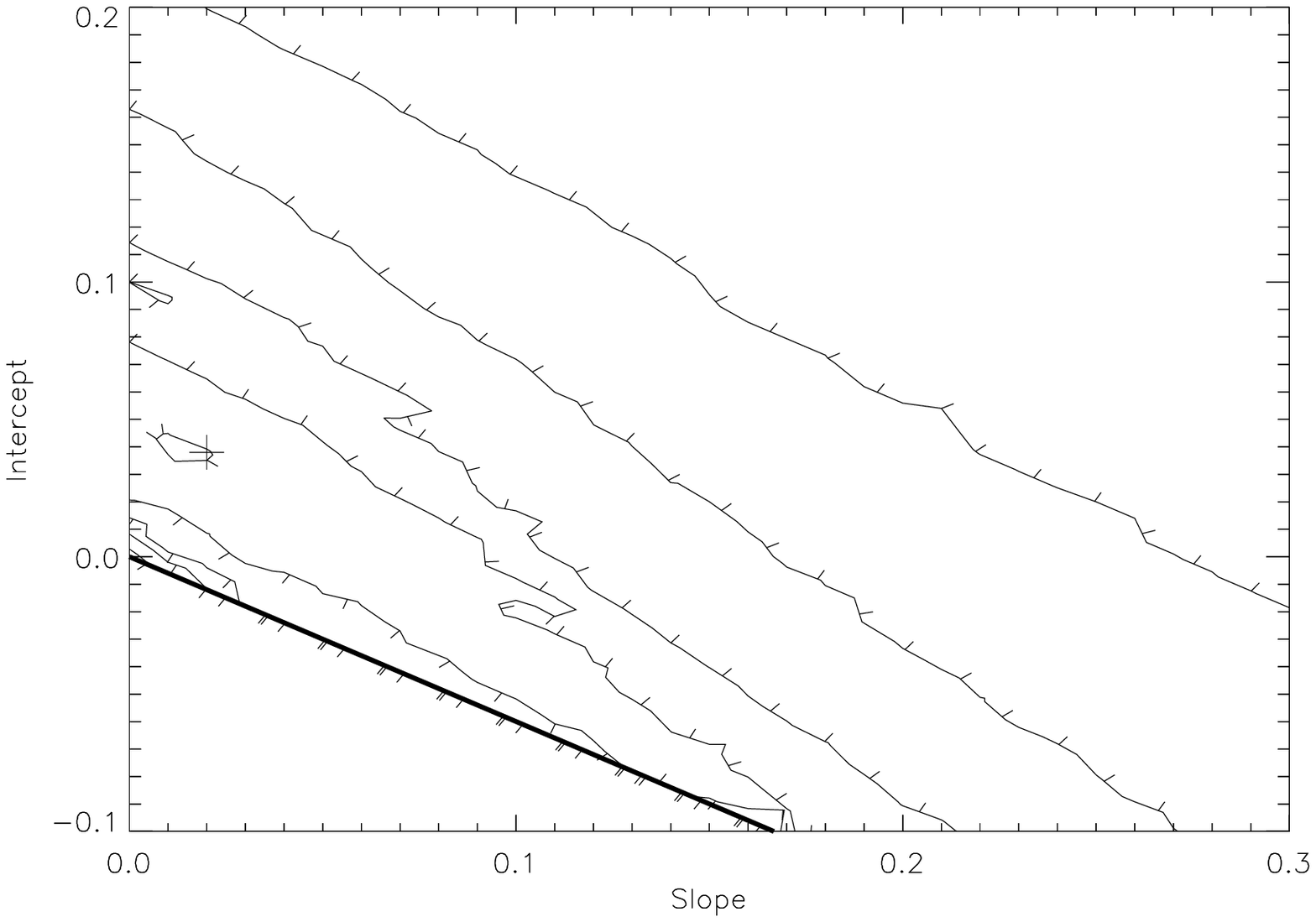} 
\figcaption[fitcont.eps]{Maximization of the
KS2D statistic (\S \protect\ref{sec:KS2D}) for spectroscopic
classification. Shown are contours of uniform $d_2=11,12,12.6,13,$ and
13.4 in the $a$-$b$ (slope-intercept) plane. Tickmarks indicate the
downhill direction, and the solid thick line indicates the $b > -0.6
a$ boundary.
The maximized value of $d_2$ is 13.44; a ``+''marks its position.
\label{fig:fitcont}}

\plotone{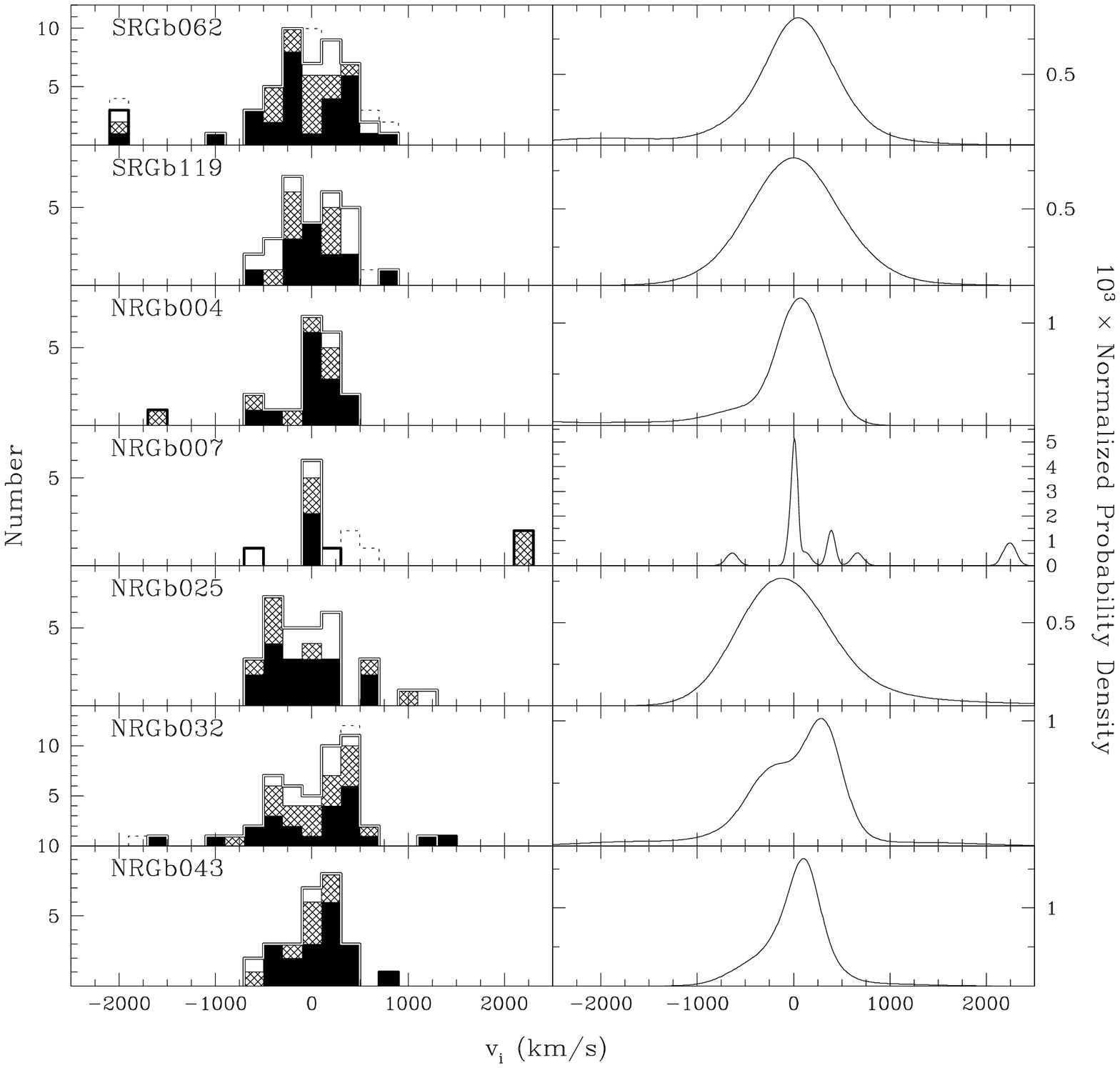}
\figcaption[hists1.eps]{The distribution of $v_i$ as defined in
section \S \protect\ref{sec:definitions}. Histograms are coded (1) by
line type: double-lined histograms show system members, heavy-lined
histograms show foreground and background galaxies inside 1.5 $h$\m\
Mpc of the system center, and dotted histograms show galaxies that are
no longer inside this region after we recenter the system; (2) by
shading: completely filled histograms show all galaxies within $0.5
h$\m\ Mpc of the system center, cross-hatched histograms those within
$0.5 h$\m--$1.0h$\m\ Mpc, and empty histograms those within
$1.0h$\m--$1.5h$\m\ Mpc. On the right is a plot of $\fka $ as defined
in equation (\protect\ref{eq:fka}).
\label{fig:hist1}}

\plotone{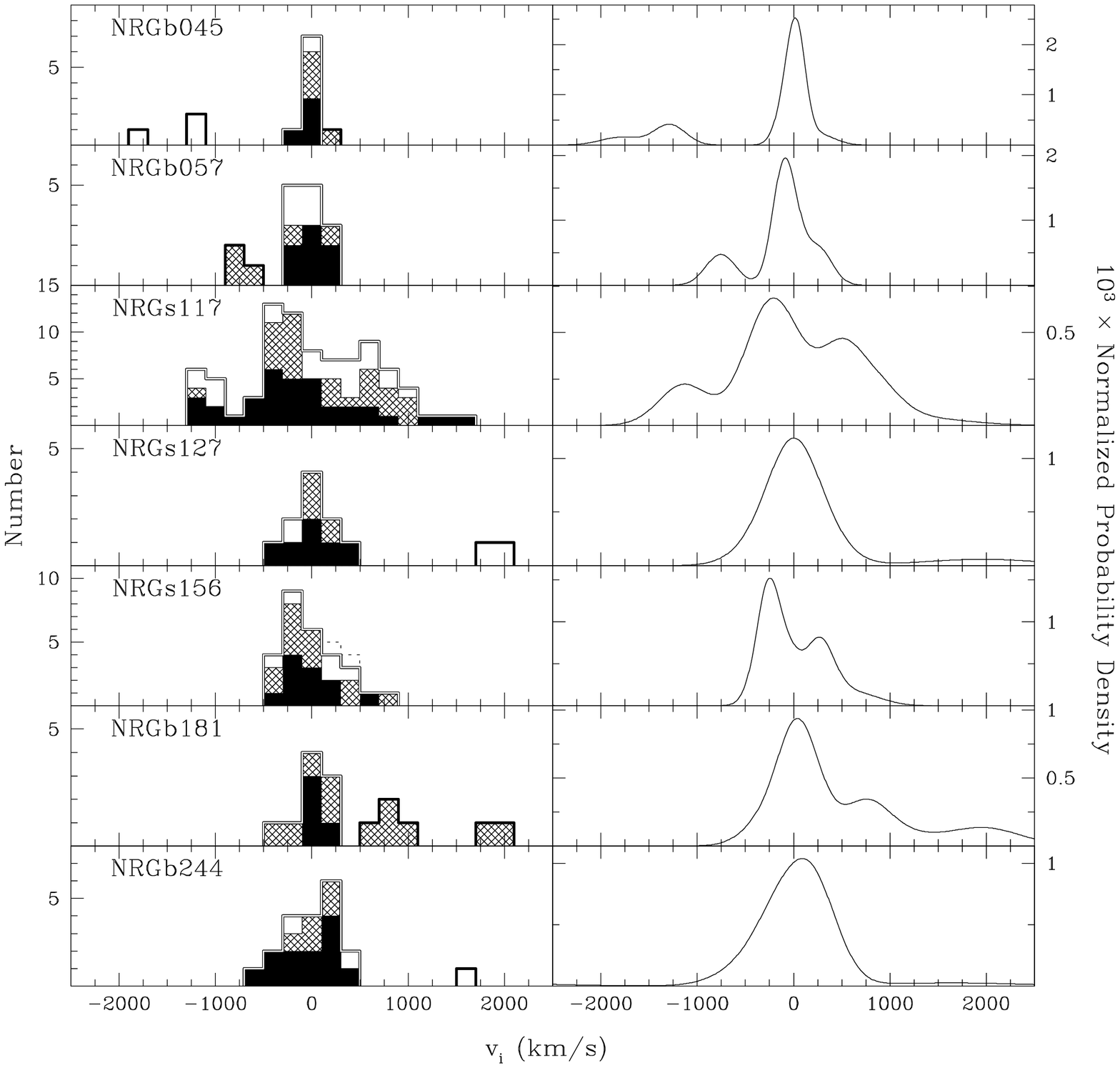}
\figcaption[hists2.eps]{See caption for Figure \protect\ref{fig:hist1}.
\label{fig:hist2}}

\plotone{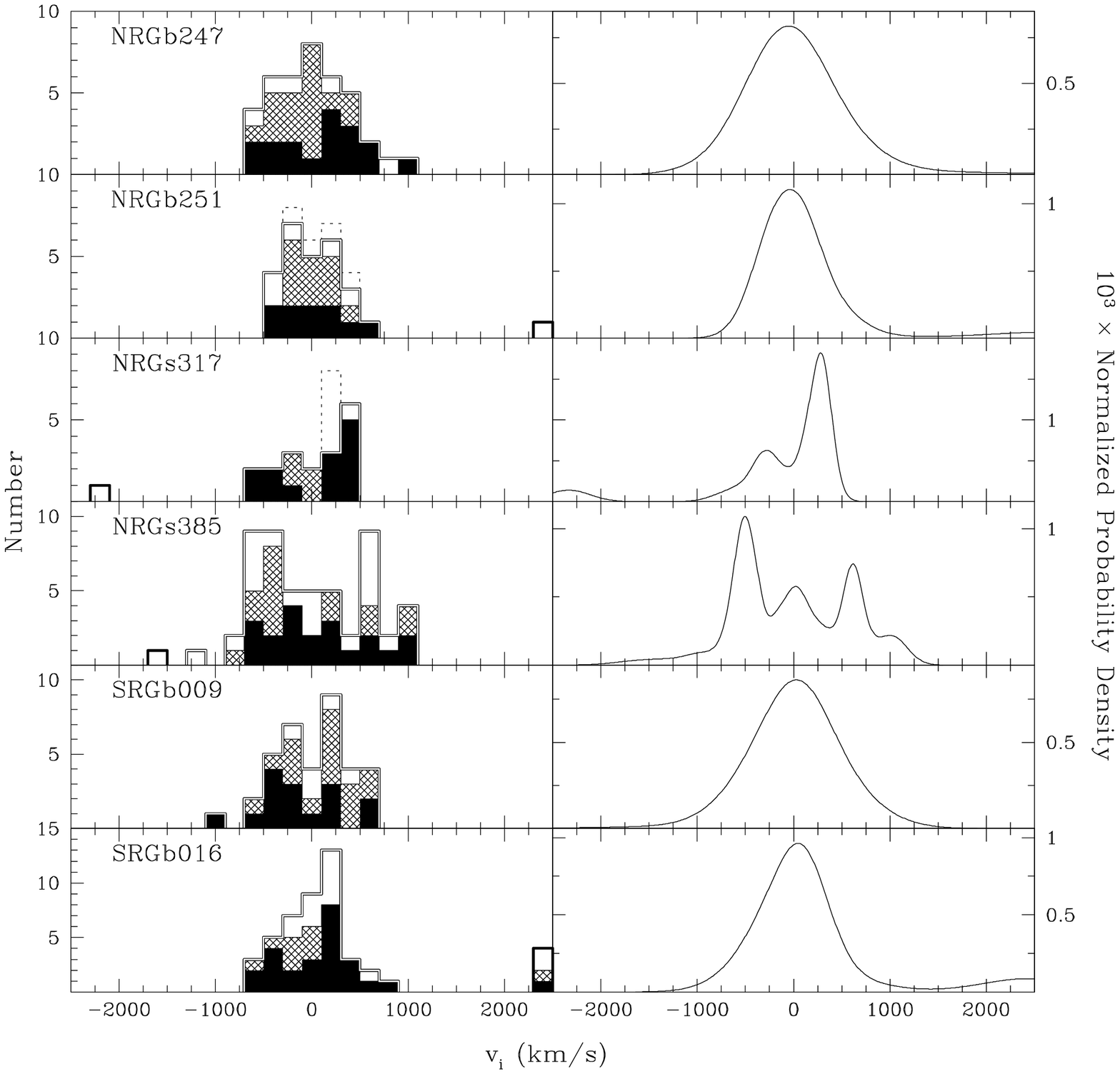}
\figcaption[hists3.eps]{See caption for Figure \protect\ref{fig:hist1}.
\label{fig:hist3}}

\plotone{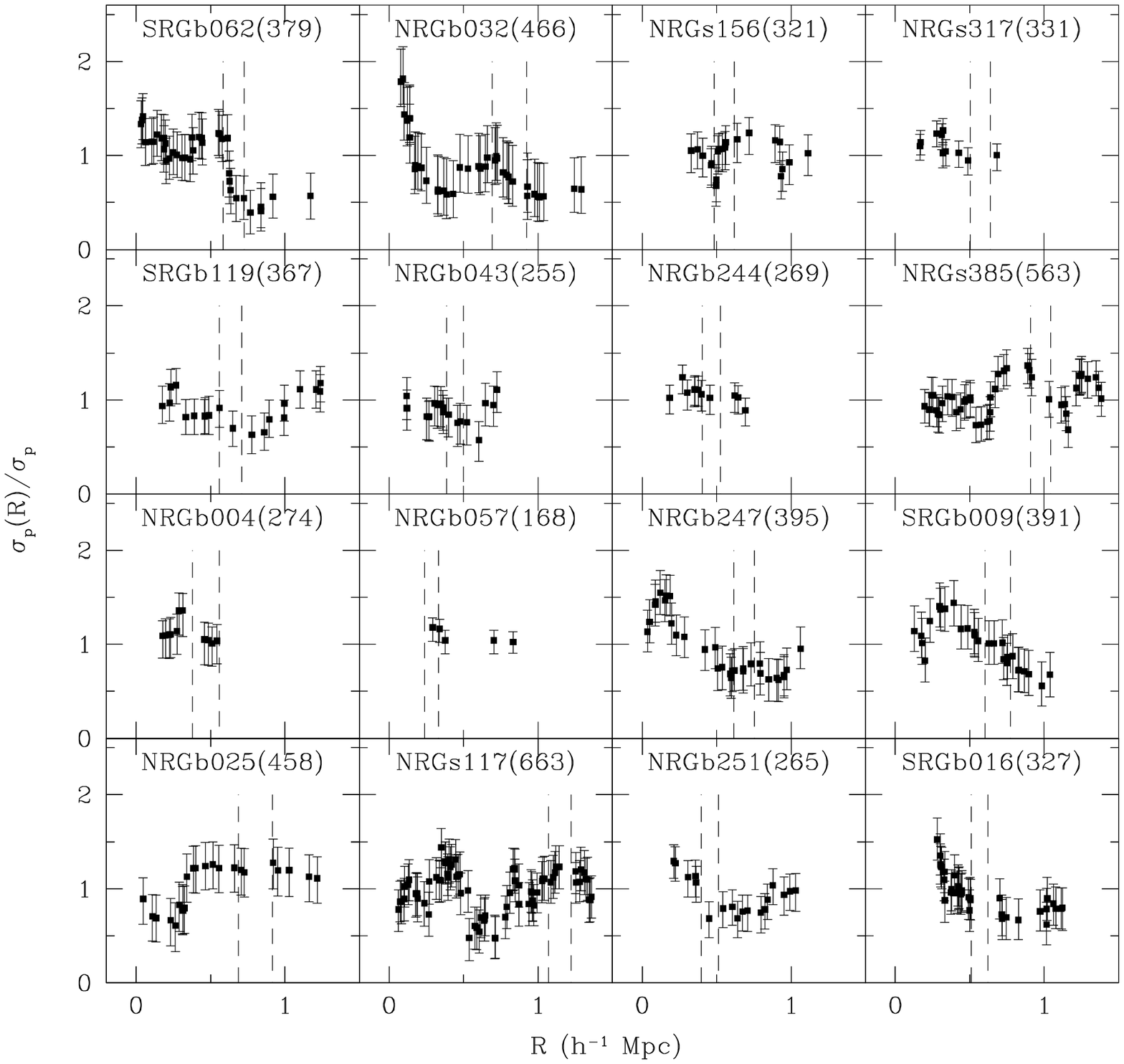} 
\figcaption[sigofr1.eps]{Velocity dispersion
profiles. The error bars show the 68.3\% bias-corrected bootstrap
confidence intervals. The number in parenthesis indicates the
global line-of-sight velocity dispersion $\sigma_p$ in km s\m.
The dashed vertical lines show the 68.3\% confidence interval
for $r_{200}$ as defined in equation (\protect\ref{eq:r200}).
\label{fig:sigofr}}

\plotone{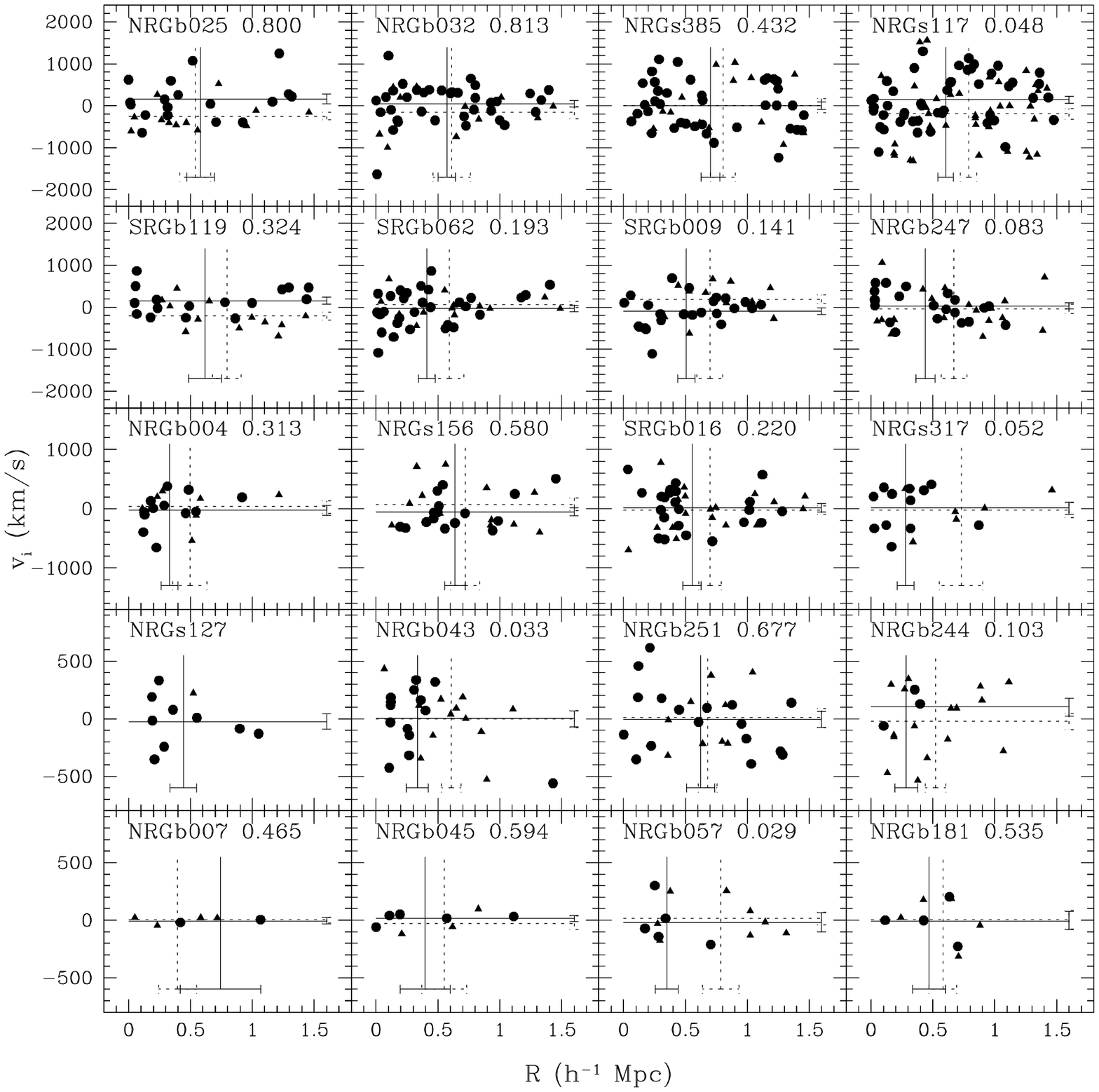} 
\figcaption[distvel.eps]{Velocity-distance
diagrams. Dotted lines represent the mean projected
distance and mean velocity of the emission-dominated galaxies
(triangles), with accompanying errors; solid lines represent these
quantities for absorption-dominated galaxies (circles). The number
next to each system name is the probability, from the Student's $t$
test, that the emission-dominated and absorption-dominated galaxies
have the same mean projected distance from the system center.
\label{fig:distvel}}

\plotone{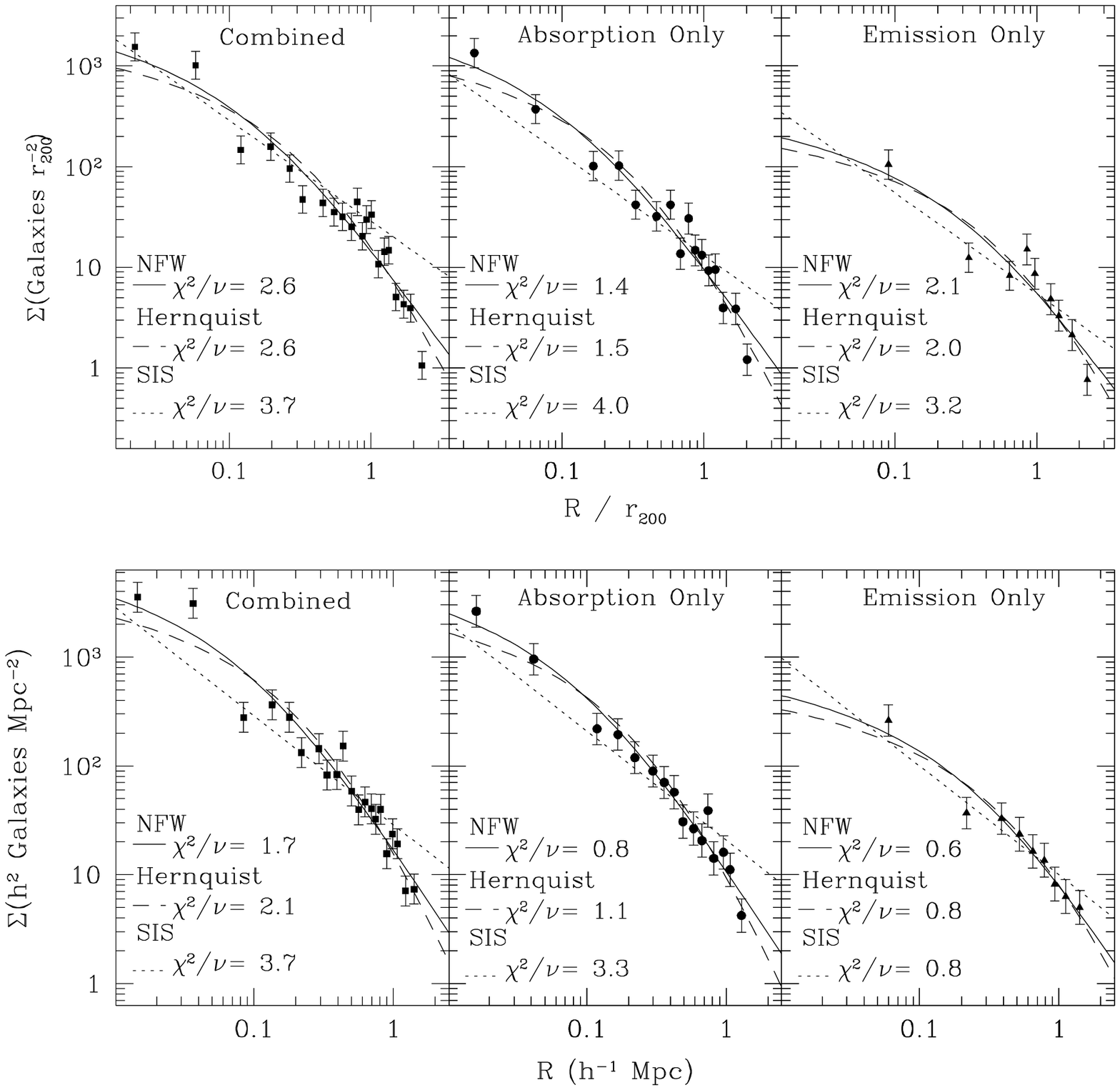} 
\figcaption[newsurface.eps]{Surface number
density profiles and best fit Hernquist (1990) NFW, and isothermal
(SIS) profiles for Sample-D. The top panels show fits with
$R$ measured in units of $\rCC$; the bottom panels show fits
with $R$ in Mpc. From left to right we have (a) the fit to the
combined emission- and absorption-dominated populations; (b) the fit
to the absorption-dominated galaxies only; and (c) the fit to the
emission-dominated galaxies only.
\label{fig:surface1}
\label{fig:firstfit}}

\plotone{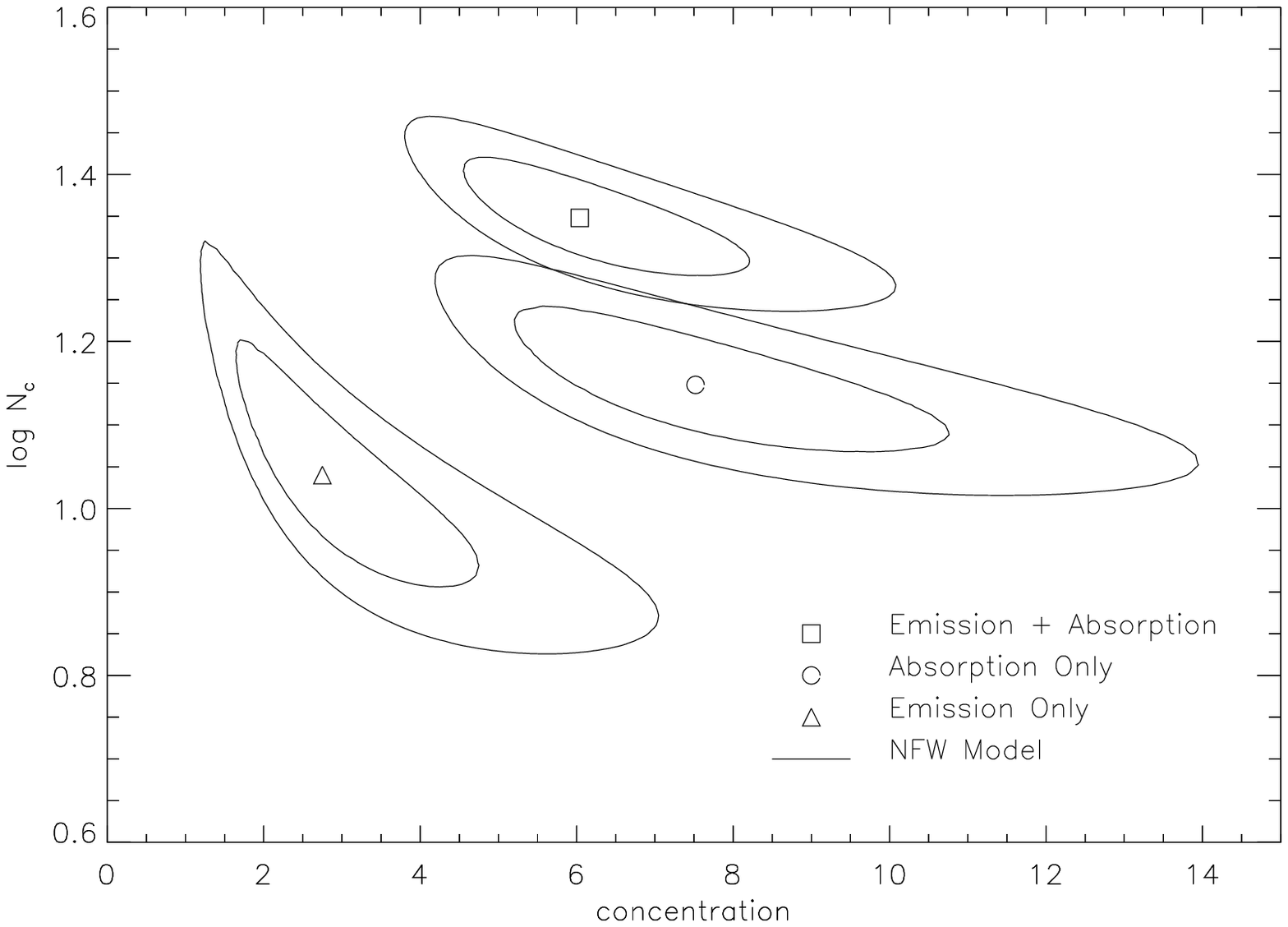}
\figcaption[r200cont.eps]{Contours of the 68.3\% and 95.4\%
confidence regions for the fits in the top panel of figure
\protect\ref{fig:surface1}.
\label{fig:confcont1}}

\plotone{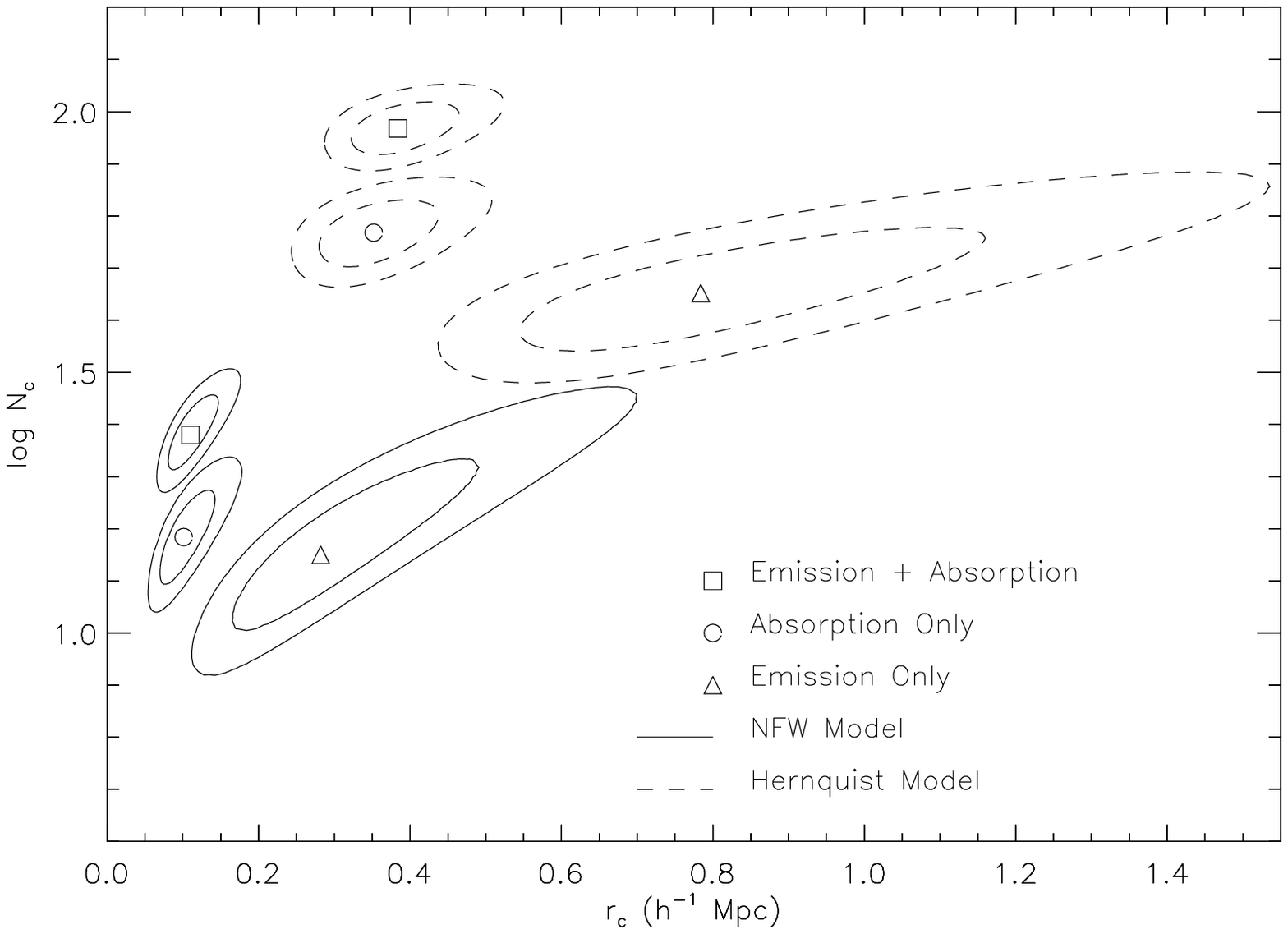}
\figcaption[confcont2.eps]{Contours of the 68.3\% and 95.4\%
confidence regions for the fits in the bottom panel of 
figure \protect\ref{fig:surface1}.
\label{fig:confcont2} \label{fig:lastfit}}

\plotone{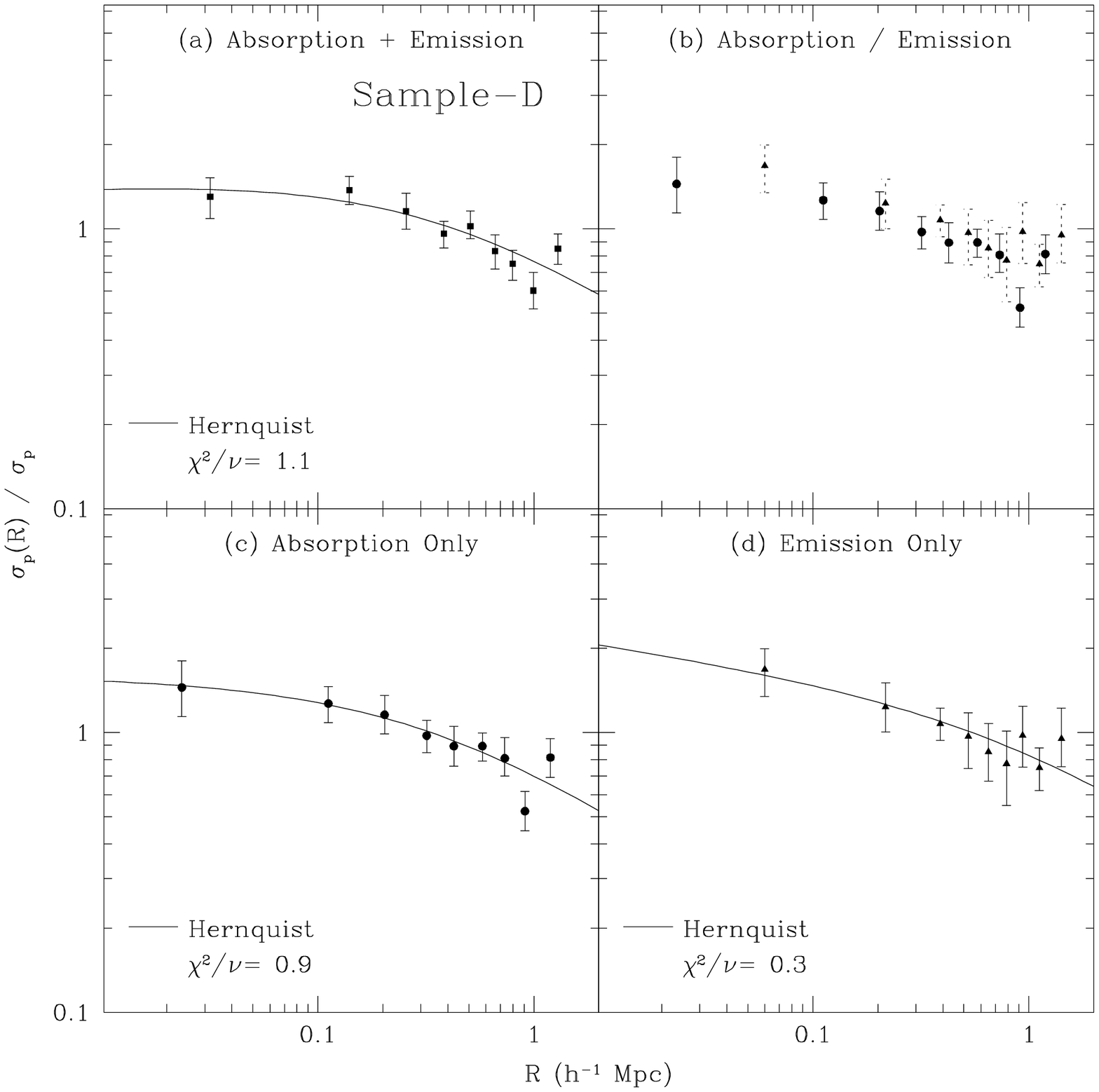}
\figcaption[velofr.eps]{Fits of the line-of-sight
velocity dispersion profile $\sigma_p(R)$ to the Hernquist (1990)
model with constant $\beta$. Shown are (a) the fit to the combined
emission- and absorption-dominated populations; (b) the two
populations separately, where solid error bars represent the
absorption- and dotted error bars represent the emission-dominated
members; (c) the fit to the absorption-dominated galaxies only; (d)
the fit to the emission-dominated galaxies only. \label{fig:velofr}}

\plotone{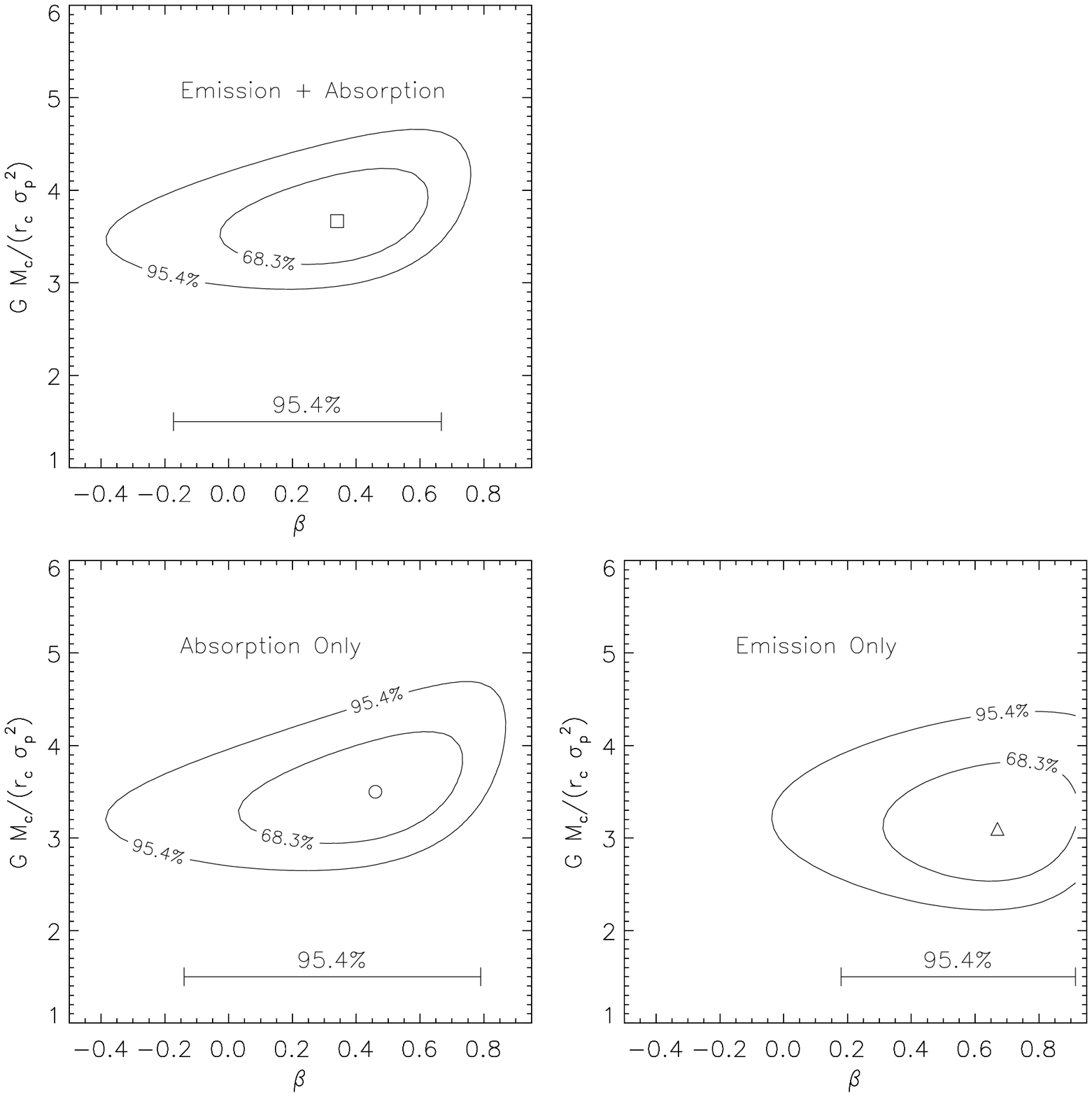} 
\figcaption[velofr.eps]{Confidence contours
corresponding to the fits in figure \protect\ref{fig:velofr}.
\label{fig:velcont}}

\plotone{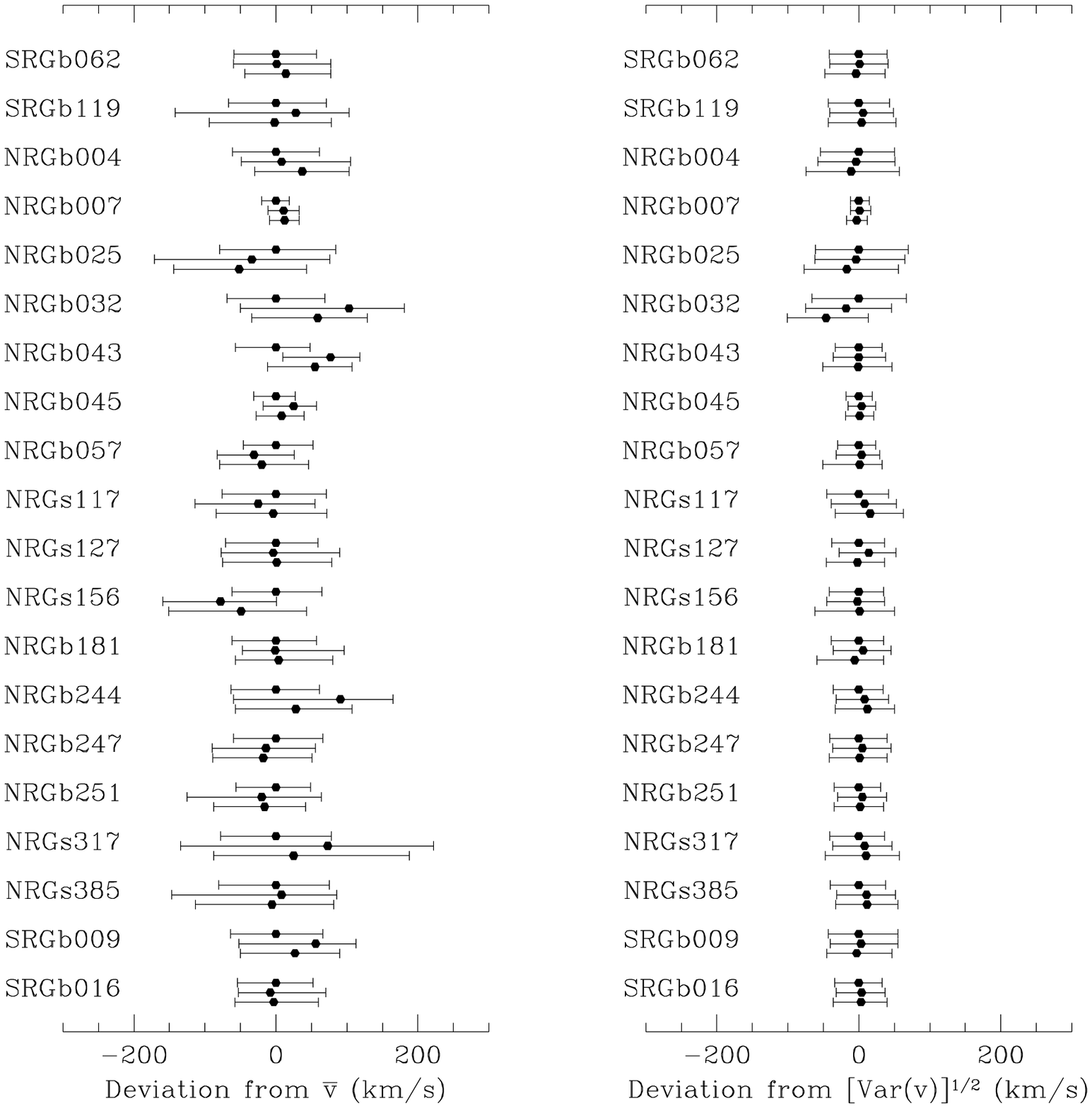}
\figcaption[robust.eps]{On the left for each system in our sample, we
plot a stack of the three estimators of location, along with their
68.3\% confidence intervals; shown are the mean (top), the median
(middle), and the biweight (bottom). We shift the values for each
system so that the mean is zero. Similarly, on the right we plot for
each system a stack consisting of the standard deviation (top), the
Gapper (middle), and the biweight (bottom) estimators of scale,
shifted so that the standard deviation is zero.
\label{fig:robust}}

\clearpage

\begin{deluxetable}{lcccccr@{$^{+}_{-}$}lr@{$^{+}_{-}$}lcc}
\tablecaption{The Deep Optical Catalog: Basic Data}
\tablehead{ \colhead{Group} & \colhead{$\alpha$} & \colhead{$\delta$} 
& \colhead{$N$} & \colhead{$N$} & \colhead{$N$} & \multicolumn{2}{c}{$\vbar$} & \multicolumn{2}{c}{$\sigma_p$} &
\colhead{} & \colhead{$P$} \nl
\colhead{ID} & \colhead{J2000} & \colhead{J2000} & \colhead{CSOC} &
\colhead{Total} & \colhead{DOC} & \multicolumn{2}{c}{km s\m} & \multicolumn{2}{c}{km s\m} & \colhead{Comments} & \colhead{point}}
\scriptsize
\startdata
SRGb062                    & 00:18:22.5 & 30:04:00 &13 & 84 & 45 & 6819 &$^{57}_{59}$&   379 &$^{ 40}_{ 42}$&  X,XC & $<0.001$ \nl 
SRGb119\tablenotemark{a}   & 01:56:13.8 & 05:35:12 & 8 & 54 & 28 & 5466 &$^{71}_{67}$&   367 &$^{ 43}_{ 43}$&  X,XC & $<0.001$ \nl 
NRGb004                    & 08:38:07.3 & 24:58:02 & 9 & 36 & 19 & 8559 &$^{61}_{61}$&   274 &$^{ 50}_{ 54}$&  X,XC &   0.587  \nl 
NRGb007                    & 08:50:29.9 & 36:29:13 & 6 & 39 &  6 & 7541 &$^{19}_{20}$&    27 &$^{ 15}_{ 12}$&   OC  & \nodata  \nl 
NRGb025\tablenotemark{b,c} & 09:13:37.3 & 29:59:58 & 5 & 35 & 31 & 6735 &$^{84}_{79}$&   458 &$^{ 70}_{ 61}$&    X,XC & $<0.001$ \nl 
NRGb032\tablenotemark{b,d} & 09:19:46.9 & 33:45:00 & 5 & 67 & 47 & 6773 &$^{69}_{69}$&   466 &$^{ 67}_{ 66}$&  X,XC & $<0.001$ \nl 
NRGb043\tablenotemark{b}   & 09:28:16.2 & 29:58:08 & 5 & 30 & 26 & 7891 &$^{48}_{57}$&   255 &$^{ 33}_{ 33}$&    OC & \nodata  \nl 
NRGb045\tablenotemark{b}   & 09:33:25.6 & 34:02:52 & 5 & 26 &  8 & 8181 &$^{27}_{31}$&    70 &$^{ 19}_{ 18}$&  X,XC & $<0.001$ \nl 
NRGb057\tablenotemark{b}   & 09:42:23.2 & 36:06:37 & 5 & 24 & 13 & 6766 &$^{52}_{46}$&   168 &$^{ 24}_{ 30}$&  X,XC &   0.684  \nl 
NRGs117\tablenotemark{b,e} & 11:10:42.9 & 28:41:38 &14 & 89 & 84 & 9799 &$^{71}_{76}$&   663 &$^{ 42}_{ 45}$&X,XC   & $<0.001$ \nl 
NRGs127\tablenotemark{b}   & 11:21:34.2 & 34:15:31 & 8 & 12 & 10 &10485 &$^{59}_{71}$&   206 &$^{ 36}_{ 38}$&    OC & \nodata  \nl 
NRGs156\tablenotemark{b}   & 11:44:54.5 & 33:15:17 & 7 & 35 & 28 & 9640 &$^{65}_{62}$&   321 &$^{ 35}_{ 42}$&X,OC   & 0.111/0.210 \tablenotemark{f} \nl 
NRGb181\tablenotemark{b}   & 12:07:35.5 & 31:26:32 & 6 & 18 &  9 & 6797 &$^{57}_{62}$&   176 &$^{ 35}_{ 39}$&    OC & \nodata  \nl 
NRGb244\tablenotemark{a}   & 13:23:57.9 & 14:02:37 & 7 & 32 & 19 & 6948 &$^{61}_{63}$&   269 &$^{ 34}_{ 36}$&  X,XC &   0.003  \nl 
NRGb247                    & 13:29:25.7 & 11:45:21 &12 & 66 & 39 & 6851 &$^{66}_{60}$&   395 &$^{ 40}_{ 41}$&  X,XC & $<0.001$  \nl 
NRGb251                    & 13:34:25.3 & 34:41:25 & 5 & 48 & 26 & 7346 &$^{49}_{56}$&   265 &$^{ 31}_{ 35}$&  X,XC & $<0.001$  \nl 
NRGs317                    & 14:47:05.3 & 13:39:46 &11 & 46 & 18 & 8873 &$^{78}_{78}$&   331 &$^{ 36}_{ 41}$&X,OC   &   0.055   \nl 
NRGs385                    & 16:17:43.9 & 34:58:00 & 8 & 61 & 53 & 9352 &$^{75}_{80}$&   563 &$^{ 37}_{ 40}$&X,XC   & $<0.001$  \nl
SRGb009                    & 22:14:46.0 & 13:50:30 & 8 & 46 & 36 & 7744 &$^{66}_{64}$&   391 &$^{ 55}_{ 43}$&  X,XC & $<0.001$  \nl 
SRGb016                    & 22:58:45.9 & 26:00:05 & 6 & 58 & 43 & 7395 &$^{52}_{54}$&   327 &$^{ 33}_{ 34}$&  X,OC & $<0.001$  \nl 
\enddata
\label{tbl:deepcatalog}
\footnotesize \tablecomments{ $N_{\mathrm CSOC}$ is the number of
galaxies classified as members in the CSOC; $N_{\mathrm Total}$ is the
total number of galaxies to $m_R \approx 15.4$ in the field of the
system; $N_{\mathrm DOC}$ is the total number of galaxies classified
as members in this work. $P_{\mathrm point}$ is the probability that
the x-ray emission in the field is due to a single point
source. Letters in the ``comments'' column denote the following. X:
has x-ray emission. XC: Position quoted is the x-ray centroid.  OC:
Position quoted is the mean optical position. }

\tablenotetext{ a } { Also studied by  Zabludoff \& Mulchaey (1998). }
\tablenotetext{ b } { Also studied by Ramella \etal (1995). }
\tablenotetext{ c } { Contains Hickson Compact Group 37 (Hickson 1982). }
\tablenotetext{ d } { Abell 779 (Abell 1958). }
\tablenotetext{ e } { Abell 1185 (Abell 1958). }
\tablenotetext{ f } { Probabilities
for the south and the north x-ray emission peaks coincident 
with member galaxies, respectively. }

\end{deluxetable}

\begin{deluxetable}{cccr}
\tablecaption{Galaxy Positions and Measured Velocities}
\tablehead{ \colhead{Galaxy ID}       & \colhead{$\alpha_{2000}$}
	&   \colhead{$\delta_{2000}$} & \colhead{$c z_i$}}
\footnotesize
\startdata
     SRGb062.01 & 00:11:57.0 & +29:29:08 &    28038  \nl
     SRGb062.02 & 00:12:11.8 & +29:19:09 &     7728  \nl
     SRGb062.03 & 00:12:17.5 & +29:52:17 &     6921  \nl
     SRGb062.04 & 00:12:28.4 & +29:32:38 &     6862  \nl
     SRGb062.05 & 00:12:38.3 & +30:06:08 &     6791  \nl
     SRGb062.06 & 00:12:45.0 & +29:22:15 &    10419  \nl
     SRGb062.07 & 00:13:12.7 & +31:08:43 &    14465  \nl
     SRGb062.08 & 00:13:45.1 & +30:11:40 &     7109  \nl
     SRGb062.09 & 00:13:55.9 & +28:45:47 &     6908  \nl
     SRGb062.10 & 00:13:57.2 & +30:52:55 &     4782  \nl
     SRGb062.11 & 00:14:01.9 & +29:25:57 &     7055  \nl
     SRGb062.12 & 00:14:13.7 & +28:52:36 &     7333  \nl
     SRGb062.13 & 00:14:55.2 & +31:06:12 &    24282  \nl
     SRGb062.14 & 00:15:07.6 & +28:52:29 &    38145  \nl
     SRGb062.15 & 00:15:15.6 & +29:21:44 &     7046  \nl
     SRGb062.16 & 00:15:22.2 & +29:39:42 &     6790  \nl
     SRGb062.17 & 00:15:23.1 & +30:43:14 &    14292  \nl
     SRGb062.18 & 00:15:24.7 & +29:38:48 &    21460  \nl
     SRGb062.19 & 00:15:28.8 & +30:43:23 &    13890  \nl
     SRGb062.20 & 00:15:43.9 & +29:39:58 &     6642  \nl
     \nodata    & \nodata    & \nodata   &  \nodata  \nl
\enddata
\label{tbl:data}
\tablecomments{This is a sample listing; the full table is available
electronically. All velocities are in km s\m; the typical uncertainty
is 40 km s\m.}
\end{deluxetable}

\begin{deluxetable}{rr@{\ }r@{\ }r@{\ }r@{\ }r@{\ }r@{\ }r@{\ }r@{\ }r@{\ }r@{\ }r@{\ }r@{\ }r@{\ }r@{\ }r@{\ }r@{\ }}
\tablecaption{Kolmogorov-Smirnov Tests for $ \sigma(R) $}
\scriptsize
\tablehead{\colhead{}&\colhead{(01)}&\colhead{(02)}&\colhead{(03)}&\colhead{(04)}&\colhead{(05)}&\colhead{(06)}&\colhead{(07)}&\colhead{(08)}&\colhead{(09)}&\colhead{(10)}&\colhead{(11)}&\colhead{(12)}&\colhead{(13)}&\colhead{(14)}&\colhead{(15)}}
\startdata
SRGb062(01)&\nodata&     **&      *&      *&    ***&    ***&    ***&     **&      *&      *&    ***&      *&    ***&      *&     **\nl
SRGb119(02)&     **&\nodata&    ***&     **&      *&      *&      *&      *&    ***&      *&      *&    ***&      *&      *&      *\nl
NRGb004(03)&      *&    ***&\nodata&     **&    ***&    ***&    ***&    ***&      *&     **&     **&      *&    ***&      *&    ***\nl
NRGb025(04)&      *&    ***&      *&\nodata&    ***&    ***&    ***&    ***&     **&    ***&    ***&      *&     **&    ***&   ****\nl
NRGb032(05)&     **&      *&    ***&    ***&\nodata&    ***&   ****&    ***&    ***&      *&      *&    ***&   ****&     **&     **\nl
NRGb043(06)&    ***&      *&    ***&   ****&      *&\nodata&    ***&    ***&    ***&    ***&     **&    ***&    ***&   ****&     **\nl
NRGs117(07)&      *&     **&      *&     **&    ***&    ***&\nodata&      *&    ***&    ***&      *&     **&      *&    ***&     **\nl
NRGs156(08)&    ***&      *&     **&    ***&    ***&     **&    ***&\nodata&     **&    ***&      *&     **&      *&      *&    ***\nl
NRGb244(09)&     **&    ***&      *&     **&   ****&    ***&      *&      *&\nodata&     **&     **&      *&     **&      *&     **\nl
NRGb247(10)&      *&      *&     **&    ***&      *&      *&    ***&    ***&    ***&\nodata&      *&      *&    ***&      *&      *\nl
NRGb251(11)&   ****&      *&    ***&   ****&    ***&     **&   ****&      *&     **&     **&\nodata&     **&      *&      *&      *\nl
NRGs317(12)&      *&    ***&      *&      *&    ***&    ***&      *&     **&      *&      *&     **&\nodata&    ***&      *&      *\nl
NRGs385(13)&     **&    ***&     **&      *&   ****&    ***&      *&    ***&     **&    ***&    ***&      *&\nodata&      *&    ***\nl
SRGb009(14)&      *&      *&      *&    ***&    ***&     **&     **&     **&      *&      *&     **&      *&    ***&\nodata&      *\nl
SRGb016(15)&    ***&      *&     **&   ****&    ***&      *&    ***&    ***&     **&      *&      *&      *&   ****&      *&\nodata\nl
\enddata
\footnotesize
\tablecomments{The results for $R$ measured in units of Mpc are above
and to the right of the dividing diagonal, while those for $R$ in
units of $\rCC$ are below and to the left. For each unique pair of
systems, the null hypothesis that $ \sigma(R) $ is drawn from the same
distribution is rejected (1) at less than the 95.4\% confidence level
if there is only one star; (2) at better than the 95.4\% confidence
level if there are two stars; (3) at better than the 99.0\% confidence
level if there are three stars; and (4) at better than the 99.9\%
confidence level if there are four stars.}
\label{tbl:ks}
\end{deluxetable}

\begin{deluxetable}
{l@{\ }rr@{\ }r@{$\,\pm\,$}lr@{\ }r@{$\,\pm\,$}l@{\ \ \ \ \ \ }r@{\ }r@{$\,\pm\,$}lr@{\ }r@{$\,\pm\,$}l}
\tablecaption{Comparison of Absorption- and Emission-Dominated Populations}
\footnotesize
\tablehead{ \multicolumn{2}{c}{} & \multicolumn{6}{c}{Velocity} & 
\multicolumn{6}{c}{Projected Distance}  \\
\cline{4-7} \cline{10-13} \\
\colhead{Group} & \colhead{$f_{\mathrm abs}$} & 
\multicolumn{1}{r}{$P(F)$} & 
\multicolumn{2}{l}{$\sigma_p^a/\sigma_p^e$} &
\multicolumn{1}{r}{$P(t)$} & 
\multicolumn{2}{l}{$\frac{\vbar^a - \vbar^e}{\sigma_p}$} & 
\multicolumn{1}{r}{$P(F)$} & \multicolumn{2}{l}{$\sqrt{\frac{\Var(R^a)}{\Var(R^e)}}$} & 
\multicolumn{1}{r}{$P(t)$} & \multicolumn{2}{l}{$\bar{R}^a - \bar{R}^e$} \\
\multicolumn{12}{c}{} & \multicolumn{2}{c}{$h$\m\ Mpc} }
\startdata
SRGb062$^{D}$ & 0.689 &  0.129 &1.49&0.35&  0.378 &-0.25&0.29&  0.436 &0.85&0.05&  0.193 &-0.18&0.15\nl
SRGb119 & 0.571 &  0.805 &0.94&0.22& {\bf 0.009} &{\bf1.00}&{\bf0.37}&  0.343 &1.33&0.04&  0.324 &-0.18&0.19\nl
NRGb004 & 0.632 &  0.914 &0.99&0.55&  0.682 &-0.21&0.55&  0.201 &0.65&0.23&  0.313 &-0.16&0.16\nl
NRGb007 & 0.333 &  0.727 &0.52&0.87&  0.629 &-0.37&1.51&  0.466 &1.49&1.55&  0.465 &0.35&0.35\nl
NRGb025 & 0.613 &  0.264 &1.39&0.46& {\bf 0.009} &{\bf0.90}&{\bf0.36}&  0.720 &1.12&0.08&  0.800 &0.04&0.18\nl
NRGb032$^{D}$ & 0.766 &  0.804 &0.96&0.26&  0.231 &0.44&0.38&  0.406 &0.83&0.03&  0.813 &-0.04&0.16\nl
NRGb043 & 0.538 &  0.775 &1.09&0.32&  0.981 &0.01&0.42&  0.543 &1.20&0.55& {\bf 0.033} &{\bf-0.27}&{\bf0.13}\nl
NRGb045 & 0.625 &  0.110 &0.39&0.26&  0.587 &0.61&1.02&  0.699 &1.44&0.56&  0.594 &-0.15&0.25\nl
NRGb057 & 0.385 &  0.636 &1.20&0.64&  0.746 &-0.21&0.62&  0.199 &0.48&0.24& {\bf 0.029} &{\bf-0.40}&{\bf0.18}\nl
NRGs117 & 0.560 & {\bf 0.069} &{\bf0.75}&{\bf0.10}& {\bf 0.036} &{\bf0.49}&{\bf0.24}&  0.725 &1.06&0.01& {\bf 0.048} &{\bf-0.19}&{\bf0.09}\nl
NRGs156 & 0.571 &  0.225 &0.71&0.17&  0.339 &-0.40&0.41&  0.447 &0.81&0.06&  0.580 &-0.08&0.15\nl
NRGb181 & 0.444 &  0.841 &0.86&0.50&  0.929 &-0.07&0.68&  0.883 &1.06&0.27&  0.535 &-0.11&0.16\nl
NRGb244 & 0.158 &  0.503 &0.55&0.24&  0.328 &0.47&0.42&  0.407 &0.48&0.29&  0.103 &-0.24&0.12\nl
NRGb247$^{D}$ & 0.538 &  0.162 &0.72&0.15&  0.603 &0.18&0.34&  0.321 &0.79&0.02& {\bf 0.083} &{\bf-0.23}&{\bf0.13}\nl
NRGb251 & 0.654 &  0.931 &1.05&0.26&  0.896 &-0.05&0.42& {\bf 0.046} &{\bf2.04}&{\bf0.09}&  0.677 &-0.06&0.14\nl
NRGs317 & 0.667 &  0.935 &1.07&0.36&  0.844 &0.11&0.50& {\bf 0.096} &{\bf0.56}&{\bf0.15}& {\bf 0.052} &{\bf-0.45}&{\bf0.19}\nl
NRGs385 & 0.717 &  0.489 &0.87&0.13&  0.985 &-0.01&0.33&  0.507 &1.18&0.02&  0.432 &-0.10&0.12\nl
SRGb009$^{D}$ & 0.667 &  0.819 &0.96&0.25& {\bf 0.055} &{\bf-0.71}&{\bf0.36}&  0.785 &1.02&0.05&  0.141 &-0.14&0.12\nl
SRGb016$^{D}$ & 0.535 &  0.862 &1.04&0.21&  0.696 &0.13&0.33&  0.532 &0.87&0.04&  0.220 &-0.15&0.12\nl
\nl
     ALL & 0.602 &  0.197 &0.93&1.96&  0.129 &0.13&3.27&  0.553 &1.04&0.00& {\bf 0.000} &{\bf-0.14}&{\bf0.03}\nl
Sample-A & 0.609 &  0.519 &0.96&2.16&  0.454 &0.07&3.58&  0.546 &1.04&0.00& {\bf 0.000} &{\bf-0.13}&{\bf0.04}\nl
Sample-D & 0.646 &  0.819 &0.98&0.11&  0.799 &-0.04&0.16&  0.266 &0.89&0.01& {\bf 0.010} &{\bf-0.15}&{\bf0.06}\nl
Sample-I & 0.585 &  0.474 &0.94&2.75&  0.258 &0.14&4.66& {\bf 0.088} &{\bf1.16}&{\bf0.01}& {\bf 0.015} &{\bf-0.11}&{\bf0.05}\nl
\enddata
\tablecomments{The fraction of absorption-dominated galaxies in each
system is $f_{\mathrm abs}$. A low value of $P(t)$ indicates that two
populations have significantly different means; a low value of $P(F)$
indicates that they have significantly different variances. Sample-A
consists of all systems except NRGs117; Sample-D contains the systems
with declining line-of-sight velocity dispersion profile, marked with
a superscript ``D''; Sample-I
consists of all systems except those in Sample-D and NRGs117.}
\label{tbl:classtab}
\end{deluxetable}

\newcommand{\mdag}{\multicolumn{2}{c}{\dag}}

\begin{deluxetable}{lrrr@{--}lcrr@{--}lcrr@{--}lcc}
\tablecaption{Surface Density Profiles}
\tablehead{\colhead{Spectral} 
& \colhead{$N_{\mathrm gal}$} 
& \multicolumn{3}{c}{NFW\tablenotemark{a}}  &
& \multicolumn{3}{c}{NFW\tablenotemark{b}}  &
& \multicolumn{3}{c}{Hernquist\tablenotemark{b}} &
& \multicolumn{1}{c}{Isothermal\tablenotemark{b}} \nl
\cline{3-5} \cline{7-9} \cline {11-13} \cline{15-15}
\colhead{Class} & \multicolumn{1}{c}{} 
& \multicolumn{1}{c}{$\chi^2/\nu$} & \multicolumn{2}{c}{$c$} &
& \multicolumn{1}{c}{$\chi^2/\nu$} & \multicolumn{2}{c}{$r_c$} &
& \multicolumn{1}{c}{$\chi^2/\nu$} & \multicolumn{2}{c}{$r_c$} &
& \multicolumn{1}{c}{$\chi^2/\nu$} }
\footnotesize
\startdata
\cutinhead{Sample-D\tablenotemark{c}\ ;  $\sigma_p = $ 327--466 km s\m.}
Absorption & 135 & 1.4 & 4.9 & 12  & & 0.8  & 0.062 & 0.16  & & 1.1  & 0.26 & 0.46 & & 3.3  \nl
Emission   &  75 & 2.1 & 1.4 & 5.6 & & 0.60 & 0.14  & 0.58  & & 0.79 & 0.50 & 1.3  & & 0.79 \nl
Combined   & 210 & 2.6 & 4.3 & 9.1 & & 1.7  & 0.074 & 0.16  & & 2.1  & 0.30 & 0.49 & & 3.7  \nl
\cutinhead{8 Groups with $\sigma_p > 350$ km s\m.}	         			      
Absorption & 232 & 2.4 & 7.6 & 16  & & 1.0  & 0.067 & 0.14  & & 2.0  & 0.30 & 0.46 & & 3.1  \nl
Emission   & 131 & 1.5 & 1.6 & 4.4 & & 1.0  & 0.29  & 0.69  & & 1.1  & 0.81 & 1.8  & & 0.9  \nl
Combined   & 363 & 2.5 & 4.3 & 7.7 & & 1.3  & 0.12  & 0.22  & & 2.4  & 0.45 & 0.65 & & 2.2  \nl
\cutinhead{\centering 9 Groups\tablenotemark{d}\ \ with $\sigma_p < 350$ km s\m.}
Absorption & 102 & 5.7 &  \mdag    & & 4.5  & \mdag         & & 3.7  &    \mdag    & & 8.8  \nl
Emission   &  92 & 3.8 &  \mdag    & & 2.5  & 0.17  & 0.59  & & 2.2  & 0.52 & 1.53 & & 3.5  \nl
Combined   & 192 & 4.2 &  \mdag    & & 3.0  & 0.12  & 0.30  & & 2.5  & 0.39 & 0.59 & & 5.1  \nl
\enddata
\label{tbl:fits}
\tablecomments{We list the 95.4\% confidence interval for each fit
parameter. $N_{\mathrm gal}$ is the total number of galaxies used in
each fit.  The core radius $r_c$ is always in units of $h$\m\ Mpc. }
\tablenotetext{a}{Fit with projected radius $R$ in units of $\rCC^{\mathrm Virial}$.}
\tablenotetext{b}{Fit with projected radius $R$ in units of Mpc.}
\tablenotetext{c}{The 5 Systems with Declining $\sigma_p(R)$.}
\tablenotetext{d}{Includes only systems with $>10$ members.}
\tablenotetext{\dag}{This fit is ruled out with high confidence ($\chi^2/\nu > 3.5$).}
\end{deluxetable}


%

\end{document}